\documentclass[twocolumn, apj]{emulateapj}

\pdfoutput=1

\usepackage{color}
\usepackage{amsmath}
\usepackage{hyperref}
\hypersetup{hidelinks} % hide hyperref link box borders (links remain active)
\usepackage{latex_style_ijc}
\usepackage{graphicx}
\usepackage{natbib}
\usepackage{subfigure}
\usepackage{multirow}
\usepackage{array}
\usepackage[table]{xcolor}
\definecolor{lightgray}{gray}{0.9}
\newcolumntype{L}{>{\centering\arraybackslash}m{3cm}}

\slugcomment{Accepted, October 17, 2013}

\shorttitle{Modeling ADI Self-Subtraction}
\shortauthors{Esposito et al.}

\begin{document}

%% LaTeX will automatically break titles if they run longer than
%% one line. However, you may use \\ to force a line break if
%% you desire.

\title{Modeling Self-Subtraction in Angular Differential Imaging: \\Application to the HD 32297 Debris Disk}

%% Use \author, \affil, and the \and command to format
%% author and affiliation information.
%% Note that \email has replaced the old \authoremail command
%% from AASTeX v4.0. You can use \email to mark an email address
%% anywhere in the paper, not just in the front matter.
%% As in the title, use \\ to force line breaks.

% One way to set author list (usually for long list).
\author{Thomas M. Esposito\altaffilmark{1}, Michael P. Fitzgerald\altaffilmark{1}, James R. Graham\altaffilmark{2}, Paul Kalas\altaffilmark{2}}
\altaffiltext{1}{Department of Physics and Astronomy, 430 Portola Plaza, University of California, Los Angeles, CA 90095-1547; esposito@astro.ucla.edu}
\altaffiltext{2}{Astronomy Department, B-20 Hearst Field Annex 3411, University of California, Berkeley, CA 94720-3411}

%% Notice that each of these authors has alternate affiliations, which
%% are identified by the \altaffilmark after each name.  Specify alternate
%% affiliation information with \altaffiltext, with one command per each
%% affiliation.

%% Mark off your abstract in the ``abstract'' environment. In the manuscript
%% style, abstract will output a Received/Accepted line after the
%% title and affiliation information. No date will appear since the author
%% does not have this information. The dates will be filled in by the
%% editorial office after submission.

\begin{abstract}

We present a new technique for forward-modeling self-subtraction of spatially extended emission in observations processed with angular differential imaging (ADI) algorithms. High-contrast direct imaging of circumstellar disks is limited by quasi-static speckle noise and ADI is commonly used to suppress those speckles. However, the application of ADI can result in self-subtraction of the disk signal due to the disk's finite spatial extent. This signal attenuation varies with radial separation and biases measurements of the disk's surface brightness, thereby compromising inferences regarding the physical processes responsible for the dust distribution. To compensate for this attenuation, we forward-model the disk structure and compute the form of the self-subtraction function at each separation. As a proof of concept, we apply our method to 1.6 and 2.2 $\mu$m Keck AO NIRC2 scattered-light observations of the HD 32297 debris disk reduced using a variant of the ``locally optimized combination of images'' (LOCI) algorithm. We are able to recover disk surface brightness that was otherwise lost to self-subtraction and produce simplified models of the brightness distribution as it appears with and without self-subtraction. From the latter models, we extract radial profiles for the disk's brightness, width, midplane position, and color that are unbiased by self-subtraction. Our analysis of these measurements indicates a break in the brightness profile power law at $r\approx110$ AU and a disk width that increases with separation from the star. We also verify disk curvature that displaces the midplane by up to 30 AU towards the northwest relative to a straight fiducial midplane.

\end{abstract}

%% Keywords should appear after the \end{abstract} command. The uncommented
%% example has been keyed in ApJ style. See the instructions to authors
%% for the journal to which you are submitting your paper to determine
%% what keyword punctuation is appropriate.


% Introduction
\section{INTRODUCTION} \label{sect:intro} 

Debris disks produced by mutual collisions of orbiting planetesimals are known to exist around several hundred nearby main-sequence stars (see, e.g., \citealt{wyatt2008} and references therein). The dust in these systems can be detected through scattered starlight or thermal emission, offering a view of the circumstellar environment during planet formation. Circumstellar debris disks represent the final stages of the planet formation process \citep{wyatt2008}. Moreover, the presence of a debris disk can be seen as an indicator of planet formation. For example, many directly-imaged extrasolar planets observed to date around main-sequence stars are located in systems with substantial debris disks (e.g., $\beta$ Pic, Fomalhaut, and HR 8799; \citealt{lagrange2010, kalas2008, marois2010_8799}). Morphological structures in the disk can act as signposts for planets that interact with the dust and planetesimals present (e.g., \citealt{mouillet1997, heap2000, wyatt2003, kalas2005_fomb, quillen2006, ertel2012, thebault2012} and references therein). These structures can also provide information about the mechanisms that replenish the dust and sculpt its distribution.

Measurements of dust-scattered light can probe the location, abundance, size, composition, and structure (i.e., porosity) of dust grains in the disk. Consequently, scattered light is sensitive to disk structure. However, measuring scattered light is difficult because although coronagraphs can be used to block light from the star, residual wavefront aberrations scatter starlight such that the faint disk signal is often overwhelmed. These aberrations are time-dependent, further complicating their removal. While adaptive optics (AO) correct for rapidly-varying atmospheric speckles, longer-timescale changes in the wave aberrations preclude simple schemes for stellar point-spread function (PSF) calibration. Fewer than 20 debris disks have been spatially resolved in scattered light since the first example was imaged in the $\beta$ Pic system \citep{smith1984}, though the majority of these systems were resolved within the last decade thanks to advances in high-contrast imaging technology and methods.

Angular differential imaging (ADI; \citealt{mueller1987, marois2006}) has proven to be an effective means for self-calibration of the time-variable residual stellar PSF. Higher contrast can be achieved by combining ADI with the method of \citet{lafreniere2007} for constructing reference PSFs through a locally optimized combination of images (LOCI). Ground-based ADI observations on an altitude/azimuth-mounted telescope require the science camera to track the telescope pupil such that, in the focal plane, the PSF orientation remains fixed but the field of view (FOV) rotates throughout the exposure sequence. For a given exposure in the sequence, one can then use other exposures to build a reference PSF that is well-suited to removing the stellar PSF. The LOCI algorithm refines construction of the reference PSF by assembling it in subsections to minimize the PSF subtraction residuals locally rather than globally. This technique has proven effective for ground-based detections of planets \citep{marois2008, marois2010_8799, lagrange2010, bonnefoy2011, currie2011, galicher2011, skemer2012, currie2012_fomb, bonnefoy2013, brandt2013, carson2013, delorme2013} and circumstellar disks \citep{thalmann2010, buenzli2010, thalmann2011, lagrange2012, milli2012, rodigas2012, thalmann2013}.

Exclusion of images nearby in time from the pool used in reference PSF construction largely mitigates the conflation of point-source PSFs with the residual stellar PSF. However, this task becomes more difficult for extended sources like disks because they subtend a larger angle and therefore require greater time separation. If the necessary images are not excluded and the extended-source PSF is blended with the residual stellar PSF, then some or all of the source's flux is removed during PSF subtraction \citep{milli2012}. For a given on-sky brightness distribution, this ``self-subtraction'' is a function of radial separation, azimuthal angle, and the ADI/LOCI parameters.

One way around the issue of self-subtraction is to use a different observational or data reduction method, though each strategy has its own drawbacks. Using a reference PSF from a disk-less star will avoid removing the disk flux but it also usually offers inferior speckle suppression because the speckle pattern changes with stellar spectral type and instrumental flexure from changes in telescope orientation. Spectral differential imaging (e.g., \citealt{smith1987, racine1999, vigan2010}) uses the invariance of the stellar PSF in simultaneous images at multiple wavelengths to differentiate it from the disk, but this method commonly requires specific spectral features in the target and instruments capable of making these observations. Polarimetric differential imaging (e.g., \citealt{kuhn2001, perrin2004, quanz2011}) separates the polarized light of the disk from the unpolarized starlight but requires an instrument with polarimetry capabilities and relatively bright targets. Recently, principal component analysis (PCA; \citealt{amara2012, quanz2013, thalmann2013}) and related algorithms like Karhunen-Lo\'{e}ve Image Projection (KLIP; \citealt{soummer2012}) have come into use as an alternative to, or in conjunction with, ADI/LOCI. These algorithms are effective and provide a nice complement to ADI/LOCI, but they can also cause self-subtraction if the set of reference images is not carefully selected to omit modes equal to those of the target object in the region being optimized.

It is also possible to reduce self-subtraction by tuning the parameters of the ADI/LOCI algorithm to perform a ``conservative'' PSF subtraction \citep{thalmann2010, buenzli2010, thalmann2011, rodigas2012, currie2012_32297, boccaletti2012, lagrange2012, thalmann2013}. These tempered implementations often have the downside of poorer noise attenuation than more aggressive formulations due to lower correlation between the reference PSF and the residual stellar PSF in the data. In addition, less-aggressive reductions do not see in as close to the star for a given ADI sequence, thus limiting investigation of the system's inner regions where planets are most likely to reside. Efforts have also been made to systematically characterize the biases introduced in brightness distributions by ADI/LOCI processing and adapt the algorithms to minimize those biases \citep{marois2010_sosie, milli2012}.

In this paper, we present a different approach to reducing the effects of self-subtraction with a new technique for forward-modeling the amount of self-subtraction of extended emission in ADI-processed images. We test the effectiveness and validity of our technique on Keck AO NIRC2 imaging of the HD 32297 debris disk at 1.6 and 2.2 $\mu$m. HD 32297 is an A star located at a distance of $112^{+15}_{-12}$ pc \citep{perryman1997}. Its circumstellar disk has been observed at wavelengths ranging from the optical through the millimeter regime and displays interesting morphological features, including brightness asymmetries and a warp. This disk is a useful test case because it is bright and has reference-star-subtracted \emph{Hubble Space Telescope} (\emph{HST}) imaging at similar wavelengths (albeit at lower spatial resolution) to our own observations \citep{schneider2005, debes2009}. In addition, the wealth of previous observations gives us many points of comparison for our work \citep{kalas2005_32297, fitzgerald2007, moerchen2007, redfield2007, maness2008, mawet2009, boccaletti2012, currie2012_32297, donaldson2013}.

We describe our self-subtraction modeling routine in Section \ref{sect:selfsubmodel}. We present details of our observations, our data reduction methods, and the results of our application of self-subtraction modeling to the HD 32297 debris disk in Section \ref{sect:applications}. Finally, we discuss the fidelity and robustness of our modeling and the implications of our results in Section \ref{sect:discussion}, and then summarize our conclusions in Section \ref{sect:conclusions}.

% Self-subtraction modeling method section
\section{A NEW TECHNIQUE FOR MODELING ADI SELF-SUBTRACTION} \label{sect:selfsubmodel}

Our technique for modeling the effects of self-subtraction in ADI is based on the operations that a modified LOCI algorithm performs on the images in a dataset to subtract the stellar PSF and speckle noise. We can use the technique to inspect where self-subtraction effects have the greatest impact. The method can also be used as a tool for comparing the LOCI-processed observations with models of the disk surface brightness derived from three-dimensional dust distribution models.

Here, we derive the self-subtraction model for the case where the data are acquired with angular differential imaging. For simplicity, we assume that each exposure in the dataset is short enough that no blurring occurs due to field rotation, though this effect could be included with a simple modification to the algorithm. Additionally, we do not account for off-axis coronagraphic PSF variation.

The ADI target exposures from which the stellar PSF is to be subtracted compose the sequence $T_i$. Let $g(r,\phi)$ be the scene on the sky (e.g., the disk), which depends on the distance from the host star $r$ and the position angle (PA) $\phi$ as measured counterclockwise from north in the sky reference frame. Actual images will also contain contributions from the star due to imperfect on-axis PSF suppression, but because the subtraction algorithm's operations are linear, we can consider the effects of the star on the rest of the scene independently. Therefore, we can define $T_i$ in terms of the scene $g$ as

\begin{equation}
T_i(r,\theta)=g(r,\theta - \theta_i)=g(r,\theta)\otimes\delta(\theta_{i}),
\end{equation}

\noindent where $\theta$ is the angle measured counterclockwise from north in the detector reference frame (i.e., the detector vertical) and $\theta_i$ is the PA of the image $T_i$ in the detector frame. In the following steps, we will define $T_i$ using the first equality. We only include the second equality to illustrate that assuming infinitesimal exposure times allows us to express $T_i$ as the scene convolved with a delta function located at $\theta_i$.

\begin{figure*}[ht]
\centering
\epsscale{1.0}
\plotone{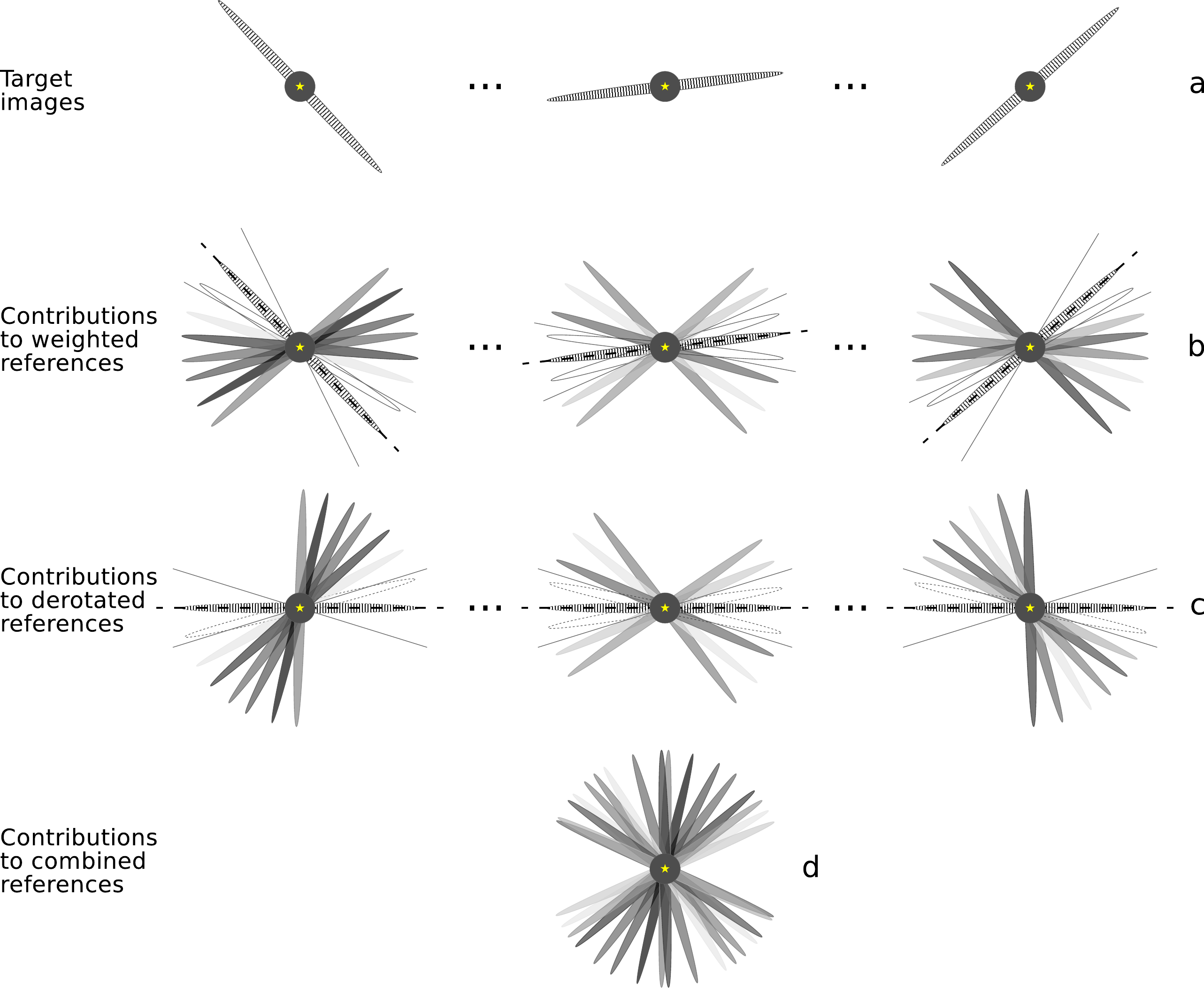}
\caption{Cartoon of the reference image selection and combination process used in LOCI and our self-subtraction model. Ellipses represent edge-on disks arranged at various PAs according to the parallactic rotation in a synthetic dataset, while the dark gray circle is centered on the star and masks the region inside of a coronagraph's inner working angle. The gray level of an ellipse indicates the weight of that image in the reference PSF combination as determined by the $c_{ij}(r)$ coefficient (darker means more heavily weighted). In row (a), the disks of the target images $T_i(r,\theta)$ are positioned at their associated $\theta_i$. In row (b), the disks of the reference images $T_j(r,\theta)$ are weighted by their $c_{ij}$ coefficients and summed to create the reference PSF for $T_i$. The disk of the target image and disks located within the minimum rotation threshold for a given separation $r$ (marked by solid lines and white fill) have $c_{ij}(r)=0$ and do not contribute to the reference PSF. Note that the amount of PA rotation is exaggerated in this cartoon for clarity, leading to fewer disks with $c_{ij}(r)=0$ than in a typical real dataset. In row (c), all reference PSFs from row (b) are derotated so that the midplane of the target image's disk lies at $\theta=0^{\circ}$. In row (d), all reference PSFs from row (c) are aligned to the star and summed to create the self-subtraction function for the final LOCI-processed image at separation $r$.}
\label{fig:selfsub_cartoon}
\end{figure*}

The reference PSF constructed for $T_i(r,\theta)$ is a linear combination of all target images weighted by their associated LOCI coefficients $c_{ij}(r,\theta)$, which the LOCI algorithm computes when trying to minimize the residuals in each subsection. Our notation differs slightly from LOCI convention here, in that $j$ ranges over all images. This requires $c_{ij}$ to be zero when an image $T_j$ is excluded from the reference PSF for target image $T_i$, i.e., when $i=j$ or when $|\theta_j - \theta_i|$ is less than a minimum rotation threshold set by the LOCI parameters (see Figure \ref{fig:selfsub_cartoon}). In addition, $c_{ij}$ is a function of both $r$ and $\theta$ by convention, but here we force it to be a function of only $r$ by taking the median over all optimization subsections at a given radius. This modification makes the modeling more tractable because we need only divide $T_i$ into annuli and not sectors within those annuli. It follows that the reference PSF-subtracted target image is

\begin{equation}
S_i(r,\theta)=T_i(r,\theta)-\sum_{j}c_{ij}(r)T_j(r,\theta) \label{eq:Si}.
\end{equation}

We build the final image $F$ by rotating all $S_i$ to align with a common sky frame and then combining those rotated PSF-subtracted images. We use the mean here for simplicity of computation; other schemes are possible (e.g., median, weighted mean). The final image $F$ is 

\begin{align}
F(r,\phi) &= \frac{1}{N}\sum_{i}S_i(r,\phi + \theta_i) \label{eq:F1} \\
	      &= \frac{1}{N}\sum_{i}\left[T_i(r,\phi+\theta_i) - \sum_{j}c_{ij}(r)T_j(r,\phi+\theta_j)\right] \label{eq:F2} \\
	      &= g(r,\phi) - \frac{1}{N}\sum_{i,j}c_{ij}(r)g(r,\phi+\theta_j-\theta_i).		
\end{align}

\noindent We can express $g(r,\phi+\theta_j-\theta_i)$ as the convolution of $g(r,\phi)$ with a delta function located at $\theta=\phi-\Delta\theta_{ij}$, where $\Delta\theta_{ij}\equiv \theta_i-\theta_j$. Here, one could incorporate blurring due to field rotation by replacing the delta function with a function of finite width, such as a top-hat function. Our final expression for the LOCI-subtracted field $F$ then becomes

\begin{equation}
F(r,\phi)=g(r,\phi) - g(r,\phi)\otimes\sum_{i,j}c_{ij}(r)\delta(\phi-\Delta\theta_{ij}) \label{eq:F4}.
\end{equation}

The second term in Equation \eqref{eq:F4} represents the ``self-subtraction function'' $z$,

\begin{equation}
z(r,\phi)=g(r,\phi)\otimes\sum_{i,j}c_{ij}(r)\delta(\phi-\Delta\theta_{ij}). \label{eq:selfsubfunc}
\end{equation}

\noindent The self-subtraction function is the linear combination of scenes weighted and positioned in the same manner as the images in the reference PSFs at a given radius. Thus, subtracting this function from the general scene $g$ will cause self-subtraction in the final image $F$ in the same locations and with the same amplitude as in the final LOCI-processed data.

An accurate calculation of the self-subtraction function is important to the quantitative investigation of structure in debris disks. LOCI parameters are tuned to maximize S/N with respect to random errors, while the self-subtraction function attempts to quantify systematic errors. Efforts to infer disk structure from ADI data must consider both types of error.

Forward-modeling has advantages over an alternative scheme to account for self-subtraction, namely injecting a model disk at a different non-overlapping angle and reducing the augmented data. This injection introduces random noise from the raw data into the self-subtraction function. Any comparison of a self-subtracted model to LOCI-processed data will then have two random noise contributions and thus two sources of uncertainty in the surface brightness. The speckle pattern near the synthetic disk may also differ from the pattern near the real disk, complicating comparison of the two after reduction. In contrast, the forward-modeling method produces a self-subtraction function free of random noise, so that the random error in a comparison of model to data comes only from the data. Practically, it may also be difficult to find a position in the data at which to inject the model disk due to interference from diffraction spikes or artifacts caused by telescope support struts. This is not an issue when using the forward-modeling technique because no alterations are made to the data itself.

Another key advantage of our forward-modeling technique is that it only requires a single LOCI reduction of the dataset (or one reduction per $N_\delta$ value if multiple $N_\delta$ values are used, as we discuss in Section \ref{sect:disk_selfsubmodel}). This means less computational expense and greater speed than methods that require LOCI reductions of multiple model-injected datasets. This speed advantage persists when comparing our technique to another self-subtraction-correction method that subtracts a model disk from the unreduced data and then measures the residuals after reduction. \citet{thalmann2013} used this method in conjunction with PCA forward-modeling to determine self-subtraction of disk emission, thereby avoiding introduction of additional noise sources associated with model-injection. Such efforts present an interesting future avenue of investigation in comparing our LOCI-based technique with PCA-based methods, but we consider it to be outside the scope of this work.

% Self-subtraction cartoon.

Finally, we suggest that it should be straightforward to adapt the self-subtraction modeling algorithm to schemes involving linear combinations of images in ADI sequences other than the LOCI version we employ in this work. For example, the algorithm can be adapted to different image-weighting schemes and has no restrictions on the number of images in the dataset. This may prove useful as new ADI-related methods are developed for the coming generation of instruments operating behind extreme adaptive optics systems that aim to reach higher contrasts and smaller inner working angles than current
% Avoid awkward page break.
%\pagebreak
\newpage
%\vspace{3 mm}
\noindent instruments (e.g., GPI, SPHERE, SCExAO; \citealt{macintosh2008, beuzit2008, guyon2011}).

\section{MODEL APPLICATIONS TO HD 32297} \label{sect:applications}

We applied the general self-subtraction modeling process outlined in the previous section to the specific case of the HD 32297 debris disk. This allowed us to test the effectiveness of the modeling method and also derive brightness profiles for the disk that were unbiased by self-subtraction. HD 32297 was also chosen partially to investigate disagreements between brightness measurements from prior ground-based and space-based observations. Brightness profiles that reflect the true dust distribution are necessary if we are going to use them to infer physical properties of the disk.  Additionally, accurate characterization of self-subtraction in our LOCI-processed images will be critical to future comparisons of observations with models of the disk surface brightness.

Previous observations of the HD 32297 debris disk have covered a wide range of wavelengths. \citet{schneider2005} revealed a disk extending from $\sim$34$-$400 AU in radius at 1.1 $\mu$m using NICMOS aboard \emph{HST}. They modeled the disk to be $10.5^\circ\pm2.5^\circ$ from edge-on and reported a brightness asymmetry in which the southwest (SW) ansa was brighter than the northeast (NE) ansa. \citet{kalas2005_32297} made ground-based, seeing-limited $R$-band observations from Mauna Kea which detected scattered light at larger scales (580$-$1680 AU) and found a brightness asymmetry similar to that of \citet{schneider2005}. \citet{kalas2005_32297} also reported position angles of the emission midplanes to the NE and SW that diverged by $31^\circ$. To explain these asymmetries, the authors proposed a collision of the disk with a clump of interstellar material as HD 32297 moves southward through the interstellar medium.

At mid-infrared (MIR) wavelengths, \citet{moerchen2007} observed the NE lobe to be brighter than the SW at 12$\mu$m beyond a radius of 0.75$''$ with T-ReCS at Gemini South, but saw no significant asymmetry at 18 $\mu$m. This work and \citet{fitzgerald2007} (11 $\mu$m imaging with Michelle at Gemini North) suggest that there is a ring of warm submicron-sized dust grains that becomes depleted within a radius of $\sim$70 AU. Resolved 1.3 mm imaging from the Combined Array for Research in Millimeter-wave Astronomy (CARMA) conducted by \citet{maness2008} showed the same SW-NE asymmetry as the near-IR and optical studies. Such a disparity in the degree of asymmetry among different wavelengths is predicted by some planetary-induced resonance models \citep{wyatt2006}, encouraging further study of this system. In the far-infrared, \citet{donaldson2013} combined \emph{Herschel} photometry with previous measurements to investigate the spectral energy distribution (SED) of the disk. Their best-fit SED model indicated a cold outer dust ring centered around 110 AU and a warm inner disk with an inner radius of $\sim$1.1 AU.

\citet{debes2009} imaged the HD 32297 debris disk at 1.6 and 2.05 $\mu$m with NICMOS and also re-examined archival 1.1 $\mu$m data collected by \citet{schneider2005}, all of which indicated the previous SW-NE asymmetry. In addition, they reported warping of the inner disk ($<400$ AU), which they modeled as the result of an interstellar cloud sculpting the disk material (similar to \citealt{kalas2005_32297}). Using the AO system deformable mirror mapped to a 1.6-m subaperture of the Hale telescope at Palomar, \citet{mawet2009} detected a truncation of the northeast ansa at $\sim$65 AU as well as a clear SW-NE asymmetry in $K_s$-band imaging, in accordance with previous studies. Recently, \citet{boccaletti2012} presented VLT/NACO AO imaging processed with various versions of ADI in $H$ and $K_s$ that did not show any significant asymmetry but did show the NE ansa to be elongated relative to the SW. They also found midplane curvature between $\sim$65 AU and $\sim$110 AU (more significantly on the NE ansa than the SW). \citet{currie2012_32297} reduced Keck NIRC2 $K_s$ data using a ``conservative'' LOCI algorithm and detected a SW $>$ NE asymmetry at small separations ($\sim$55$-$65 AU) but their radial brightness profiles also exhibited an unmentioned NE $>$ SW asymmetry at $r\approx$ 120$-$145 AU. They reported a northward disk curvature similar to previously-mentioned works, as well. The circumstellar environment of HD 32297 is also unique because of a significant gas component detected in Na I absorption by \citet{redfield2007} and in [CII] emission by \citet{donaldson2013}, a rarity for debris disk systems.

\subsection{Observations} \label{sect:obs}

We observed HD 32297 using the Keck II AO system and a coronagraphic imaging mode of the NIRC2 camera. We took 30 images with individual exposure times of 30 s in the $H$-band filter ($\lambda_0=1.63\ \mu \mathrm{m}, \Delta\lambda=0.296\ \mu$m) on 2005 October 22, and 9 images with individual exposure times of 60 s in the $K_s$ filter ($\lambda_0=2.15\ \mu \mathrm{m}, \Delta\lambda=0.311\ \mu$m) on 2007 September 22. The camera was operated in ``narrow'' mode, with a $10''\times10''$ FOV and a pixel scale of 9.95 mas $\mathrm{pixel^{-1}}$ \citep{yelda2010}. A 200 mas radius coronagraph mask artificially eclipsed the star in $H$ science images while a 400 mas radius mask was in place for $K_s$ science images. Airmass ranged from 1.02 to 1.20 across the two nights and the AO loop was closed with HD 32297 serving as its own natural guide star. The PSF full width at half maximum (FWHM) was within 10\% of the diffraction limit in all images, with the diffraction limit at the central wavelength equal to 41 mas (4.1 pixels) in $H$ and 54 mas (5.4 pixels) in $K_s$.

We employed ADI for all science observations. Using this technique, we acquired a sequence of images with the camera rotator in vertical angle mode, in which the altitude/azimuth-mounted telescope and the rotator were adjusted such that the telescope PSF's orientation was held fixed relative to the camera and AO system optics. This caused the FOV to rotate throughout the sequence while the PSF orientation was fixed relative to the detector. Our images spanned $30.5^\circ$ of angular rotation in the $H$ dataset and $37.2^\circ$ in $K_s$.

For calibration purposes, we observed standard stars SJ 9105 \citep{persson1998} and FS 30 \citep{hawarden2001} unocculted to determine the photometric zeropoint in $H$ and $K_s$ bands, respectively. We note that our observations of SJ 9105 showed it to be a visual double, with a second source $\sim$0.5$''$ to the southeast of the standard star. Although the second object is relatively faint (total flux only $\sim$4\% that of the standard), it may have affected our calculation of the photometric zeropoint and systematically biased our photometry downward by $\sim$4\%. Flux densities used for flux conversion were taken from \citet{tokunaga2005}.

% Data reduction
\subsection{Data Reduction} \label{sect:reduction}

We used the same procedure to reduce both the $H$ and $K_s$ datasets. After bias-subtraction and flat-fielding, we masked cosmic ray hits and other bad pixels. Next, we aligned the individual exposures via cross-correlation of their stellar diffraction spikes \citep{marois2006}. Following this, radial profile subtraction suppressed the stellar halo and the sky background by effectively acting as a high-pass filter.

We applied a modified LOCI algorithm \citep{lafreniere2007} to our data to suppress the stellar PSF and quasistatic speckle noise. For each image in a dataset, LOCI constructs a unique reference PSF from an optimized linear combination of other images in the dataset. The reference is constructed in azimuthally-divided subsections of annuli centered on the star. In each subsection, the coefficients $c_{ij}$ of the linear combination are chosen so as to minimize the residuals of the PSF subtraction. To simplify our self-subtraction modeling procedure (see Section \ref{sect:disk_selfsubmodel} for further explanation), we modified the LOCI algorithm to first compute the coefficients as functions of both azimuthal angle and radius, following the prescription presented by \citet{lafreniere2007}. Then, the algorithm takes the median of each $c_{ij}$ across all subsections in a given annulus, and replaces the original coefficient with that median value. Only after this do we perform the linear combination that creates the reference PSF subsection. This reduces $c_{ij}$ to being only a function of radius and provides a single coefficient per annulus for each image, as described in Section \ref{sect:selfsubmodel}.

After using LOCI to subtract the stellar PSF from all images in the dataset, we derotated the PSF-subtracted images, and then averaged them to create our final image. Averaging, rather than median combination, was important for preserving the linearity necessary for self-subtraction modeling and for simplifying the calculation of the self-subtraction function (see Section \ref{sect:selfsubmodel}).

% Final LOCI-processed images in H and K.
\begin{figure*}[ht]
\centering
\epsscale{1.1}
\plotone{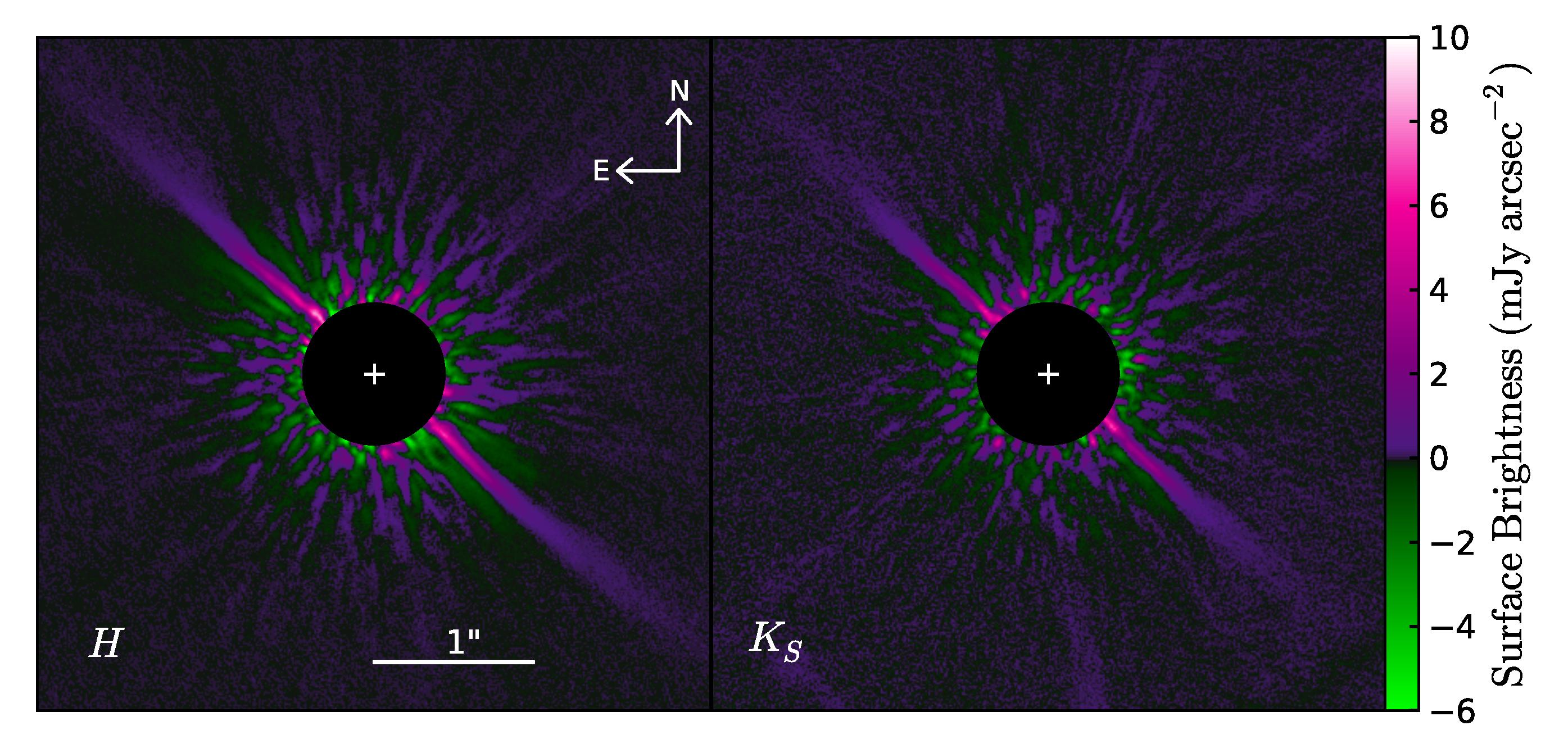}
\caption{Final LOCI-processed images of the disk surface brightness in scattered light in $H$ (\textit{left}) and $K_s$ (\textit{right}) bands. The disk is seen approximately edge-on with one ansa to the NE and the other to the SW of the star.  Green areas are negative-brightness regions created by self-subtraction or imperfect PSF subtraction. The star location is marked by a white cross. The black circle masks the innermost radii where either residual speckle noise dominates or the coronagraph mask obscures the disk.}
\label{fig:LOCI_finals}
\end{figure*}

We maximized the signal-to-noise ratio (S/N) of the final image by tuning the LOCI parameters to achieve a balance between noise attenuation and disk flux retention. The S/N used in tuning LOCI parameters was calculated by performing aperture photometry at multiple, equally-spaced radial separations between $r=21$ and $r=300$ pixels. We tuned five LOCI parameters. The azimuthal width of the disk PSF is given by $W$ (in pixels), and $N_{\delta}$ represents the minimum gap allowed (in units of $W$) between the disk midplane in the target image and the midplanes in the images used as references. The radial width of the subtraction subsections, in which the subtraction residuals are locally minimized, is set by $dr$ (in pixels). The parameters $g$ and $N_a$ are dimensionless and, along with $W$, determine the radial and azimuthal widths of the optimization subsections. See \citet{lafreniere2007} Section 4.1 for more detailed definitions of all five parameters.

For the $H$ data, we determined optimal parameter values of $W=10$ pixels ($\sim$11 AU), $N_{\delta}=0.5$, $dr=5$ pixels ($\sim$5.5 AU), $g=0.1$, and $N_a=250$. Our value for $W$ is based on an estimate of the average disk FWHM at small $r$. For the $K_s$ data, we determined slightly different optimal parameter values of $W=10$ pixels, $N_{\delta}=0.25$, $dr=5$ pixels, $g=0.1$, and $N_a=150$. Additional images were created from each dataset for use in our profile fitting routine (Section \ref{sect:disk_selfsubmodel}), with values of $N_\delta$ ranging from 0 to 5 and the other parameters set to their optimal values listed above.

% LOCI image results
\subsection{PSF-Subtracted Images}

We present the final LOCI-processed $H$- and $K_s$-band images in Figure \ref{fig:LOCI_finals}. We spatially resolve the disk in scattered light at projected separations of $\sim$50$-$300 AU ($0.45''$$-$$2.7''$). Residual speckle noise significantly contaminates the disk emission inward of 50 AU and creates confusion with the disk signal. Negative-brightness regions occur above and below the disk midplane as a result of self-subtraction by LOCI processing.

Upon visual inspection, the ansae appear largely symmetrical in brightness and size in both bands. The only apparent feature is a curvature of both ansae towards the northwest that is more pronounced at larger separations. This curvature is discussed in more detail in Sections \ref{sect:midplane} and \ref{sect:structure}.

% Self-subtraction modeling of HD 32297
\subsection{Brightness Modeling Accounting for Self-Subtraction} \label{sect:disk_selfsubmodel}

To demonstrate the effectiveness of our self-subtraction modeling algorithm, we applied it to our observations of the HD 32297 disk. By combining the framework from the previous sections with a simple parameterization of the disk's brightness distribution, we produce a model of the two-dimensional surface brightness as it appeared before undergoing LOCI self-subtraction. We then extract radial profiles for disk parameters, such as the one-dimensional integrated brightness, from this model in order to investigate the distribution of dust in the disk. This method is slightly more complex than aperture photometry, but it produces brightness measurements that are largely independent of the self-subtraction. It also avoids complicated three-dimensional radiative transfer models and data inversion. Our goal is not necessarily to produce a perfect representation of the disk, but rather to construct a model that is useful for verification of our self-subtraction modeling method and for estimating the disk's physical parameters. These estimates can later be used as starting points for constructing more complex three-dimensional models that may provide more accurate measurements of those parameters.

% Slice example plots.
\begin{figure}[ht]
\centering
\epsscale{1.2}
\plotone{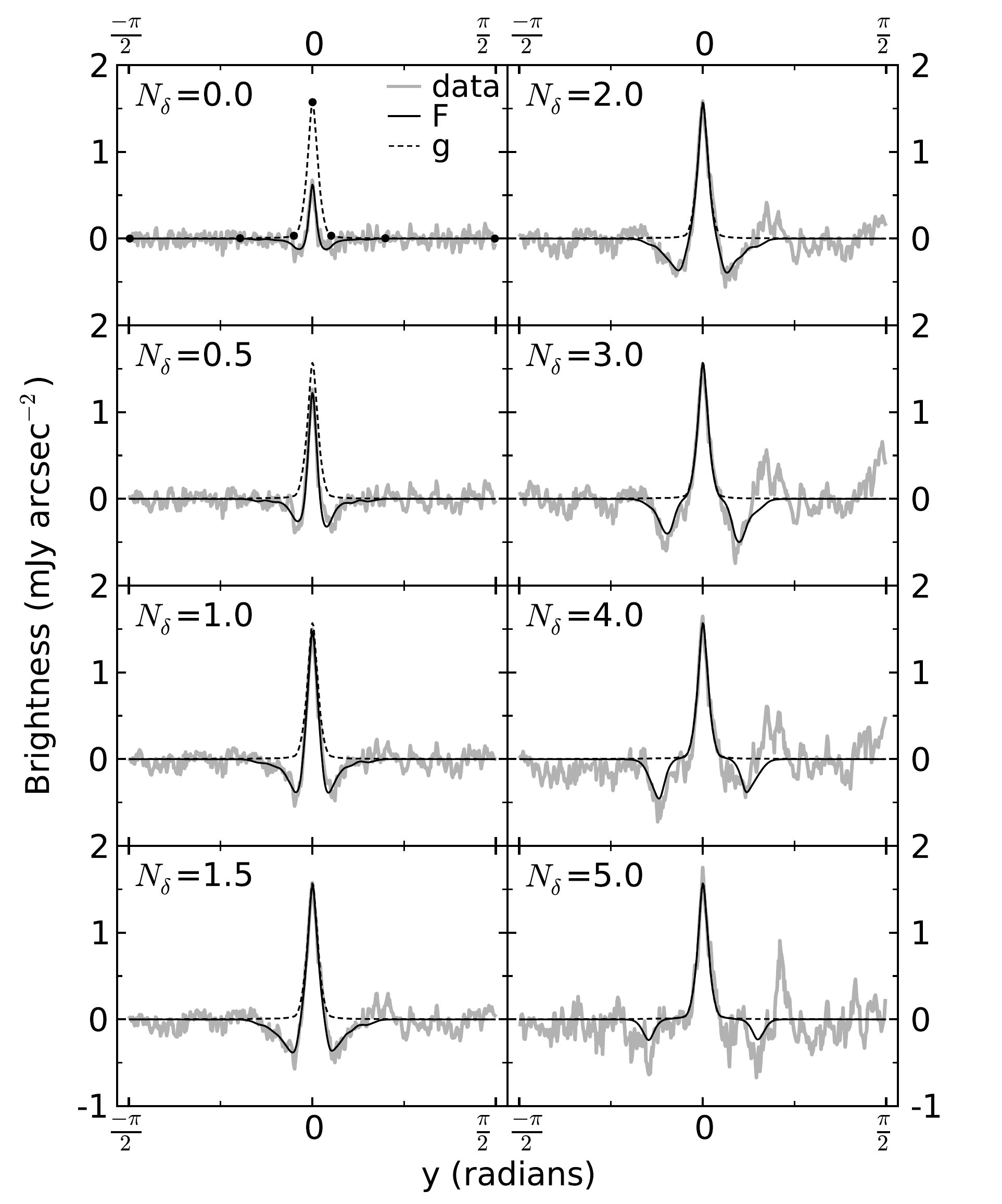}
\caption{Examples of one-dimensional semiannuli fit by our self-subtraction modeling algorithm. The different panels show semiannuli from the $H$-band images reduced using different $N_\delta$ values, but all from the SW ansa at $r=133$ AU. The thick gray lines represent the data, the solid black line is the self-subtracted model ($F$) fit to the data, and the dashed line is the best-fit cubic spline model of the underlying disk vertical brightness profile ($g$) with filled circles marking the control point locations in the first panel. The scene $g$ is constant in all panels but the self-subtracted model differs with $N_\delta$.}
\label{fig:slices}
\end{figure}

We begin by defining two coordinate systems so we can more easily extract data from our images and also construct models to compare with those data. First is a standard Cartesian coordinate system ($x$,$y$) with the star at the origin, $x$ along the horizontal axis (positive is to the right), and $y$ along the vertical axis (positive is up). Second is a polar coordinate system ($r$, $\theta$) with the star at the origin and $\theta=0^\circ$ along the positive $x$-axis. We then rotate the final LOCI-processed image counterclockwise by $42.5^\circ$ (the complement of our estimated disk PA of $\sim47.5^\circ$) so the disk midplane lies roughly along the $x$-axis, and divide the image in two along a line parallel to the $y$-axis and passing through the star location. This gives us separate SW and NE sections of the image and allows us to deal with one ansa at a time.

For one of the ansae, we further divide the corresponding image section into semicircular annuli (semiannuli) that are centered on the star, have the center of their arc aligned with $\theta=0^\circ$, and are one pixel wide in the radial direction. Each semiannulus contains the disk's one-dimensional brightness profile at $r=r_0$, where $r_0$ is equal to the semiannulus' radius of curvature. For computation purposes, we project the semiannulus onto a line, changing coordinates of the semiannuli from polar to Cartesian via the transformations $y=r\mathrm{sin}(\theta)$ and $x=r$. This makes the brightness profiles functions of vertical ($y$) and radial ($x$) distance from the star. These one-dimensional vertical profiles represent the data to which we fit our self-subtraction model. Figure \ref{fig:slices} shows example profiles from our $H$ band dataset processed with various $N_\delta$ values (with $y$ plotted along the horizontal axis).

Next, our model function for the scene $g$ (as it is defined in Section \ref{sect:selfsubmodel}) represents an estimate of the disk vertical brightness profile before it underwent self-subtraction. In this case, we approximate the shape of the disk's vertical profile using a monotonic cubic spline. For simplicity, we assume the profile is symmetric about the disk midplane and that the peak brightness occurs at the midplane. We divide the profile at the midplane and select four interpolation control points along one half of the profile. One control point is located at the peak, one is located in the wing of the profile, another is in the profile tail, and the final point is positioned at the end of the tail of the profile. The brightnesses of these control points are variables $b_0$, $b_1$, $b_2$, and $b_3$, respectively. Both $b_0$ and $b_1$ are allowed to vary in the least-squares fit. We fix $b_2 \approx b_3 \approx 0$, assuming that the disk flux is negligible far from the midplane. This number of control points sufficiently approximates the general shape of the disk's vertical profile for our proof-of-concept test without compromising computational speed. Adding more control points would likely be one way to improve upon our model and increase the accuracy of our measurements. Once values for the control points are assigned, we reflect them about the midplane, so that we can interpolate the entire profile. We do so over the range $y$ to get $g$, which we insert into a one-dimensional version of Equation \eqref{eq:F4} that replaces $r$ and $\theta$ with $x$ and $y$. The third and final fit parameter is the position of the midplane ($y_{\mathrm{mid}}$), which sets the position of the peak control point along $y$.

To obtain initial estimates for the fit parameters, we derive them from a Lorentzian profile assumed to have a peak brightness equal to the maximum brightness of the vertical profile and a FWHM drawn from a power law established from a conservative (large $N_\delta$) reduction of the data. The three parameters depend only on $x$, therefore, the function $g$ that they produce is also a function of $x$. Early tests showed that a pure Lorentzian function does not approximate the shape of the disk well at all separations, leading us to use the spline as our model function. We chose a simple model for the vertical profile because our primary objective was to verify our procedures for modeling self-subtraction in the disk, and not necessarily to model the disk itself in fine detail.

% Model image stack for H.
\begin{figure*}[ht]
\centering
\epsscale{1.0}
\plotone{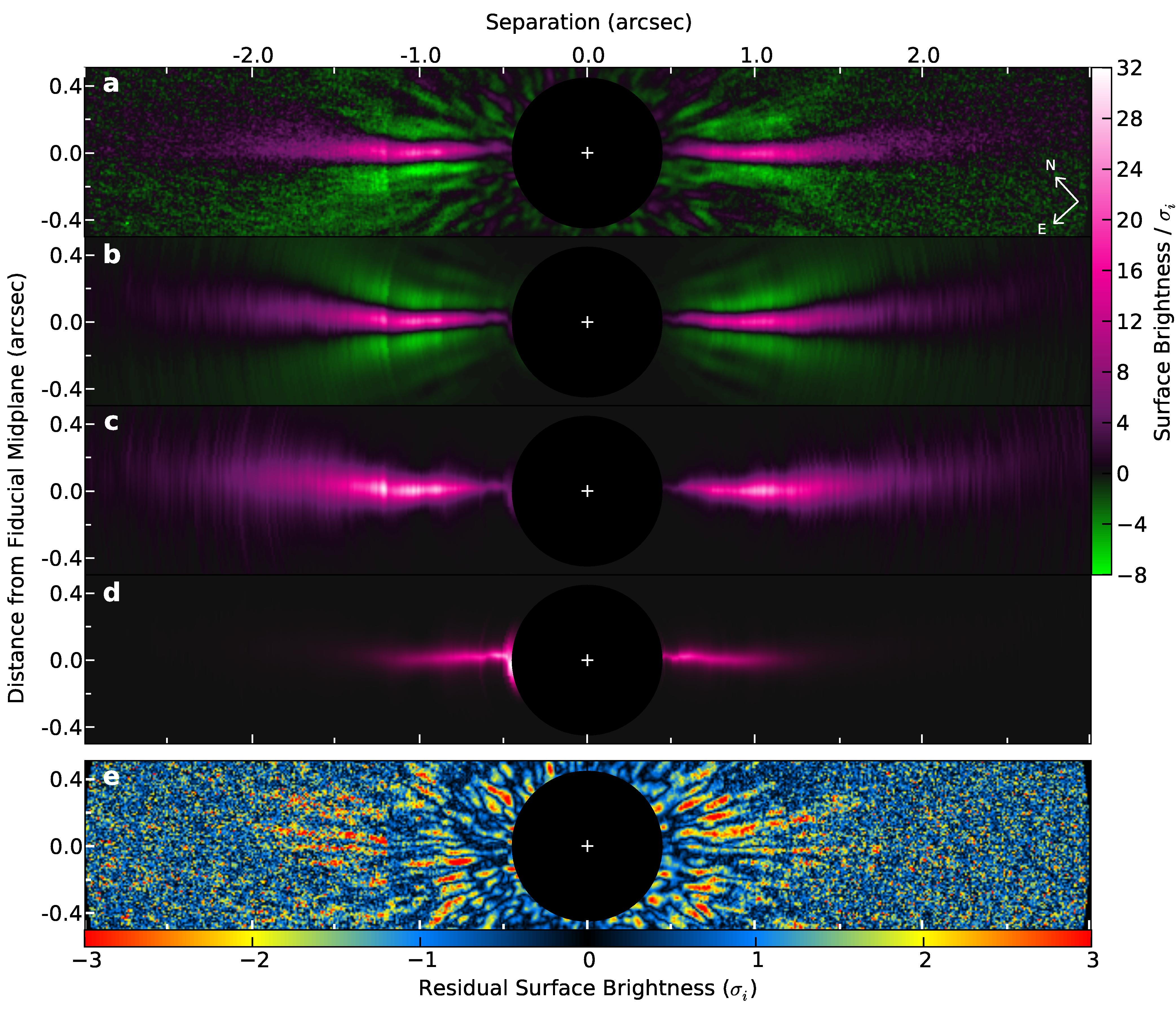}
\caption{$H$-band images of the disk surface brightness. All panels except (d) have been scaled by $\sigma_i$, which is the separation-dependent random error in the data at each pixel $i$ (not to be confused with $\sigma$ calculated from a $\chi^2_\nu$-scaled covariance matrix). (\textit{a}) LOCI-processed image from Figure \ref{fig:LOCI_finals}, derotated by $42.5^\circ$ so the fiducial midplane lies along the horizontal. (\textit{b}) Model of the self-subtracted disk produced by subtracting the forward-modeled self-subtraction function from the best-fit model of the disk's underlying brightness distribution. Negative-brightness regions above and below the midplane match those in the real data. (\textit{c}) Best-fit model of the disk's underlying brightness distribution absent of self-subtraction. Disk curvature towards the northwest is clearly visible. (\textit{d}) Same as in (\textit{c}) except the brightness is not scaled by $\sigma_i$ and follows a linear color stretch from 0$-$13 mJy arcsec$^{-2}$. (\textit{e}) Deviate map for (\textit{a}) - (\textit{c}) normalized by $\sigma_i$. Note that the deviates were not scaled by $\chi^2_\nu$ and thus do not account for correlated pixels or over-constrained model parameters.}
\label{fig:Hmodels}
\end{figure*}

As Figure \ref{fig:selfsub_cartoon} (row a) illustrates, the disk is oriented at a different PA in each image in the dataset. Data recorded at the telescope during observations provide the PAs for each image ($\theta_i$), which we transform to $y_i$ and insert into the self-subtraction function (Equation \ref{eq:selfsubfunc}). The same $c_{ij}$ coefficients used in the LOCI processing of the image being modeled are also substituted into the expression. We include an additional weighting coefficient in the summation in Equation \eqref{eq:selfsubfunc} based on a rough approximation of the Strehl ratio\footnote{As a proxy for the Strehl ratio, we used a circular aperture of radius 3 pixels to measure the flux of the star through the coronagraph mask in an image relative to the mean of this flux across all images in the dataset.} of each reference image $T_j$. This weighting attempts to account for variations between images due to seeing and it is only applied during modeling. The Strehl was fairly stable and the PSF core remained diffraction-limited across the observations, as weights for both datasets were in the range 0.9$-$1.2 (with 1.0 as the mean). As shown in Figure \ref{fig:selfsub_cartoon} (row b), for a given target image $T_i$, some $T_j$ have $c_{ij}=0$ because those reference images are excluded from the reference PSF based on insufficient field rotation (the minimum threshold being set by LOCI parameters $N_\delta\times W$). With all of the components assembled, we compute the self-subtraction function in Equation \eqref{eq:selfsubfunc}, and then insert this function into Equation \eqref{eq:F4} to produce a self-subtracted model of the vertical profile, which we call $F$.

We use a weighted least-squares routine to compare $F$ with the observed profile and determine the best-fit values for $b_0$, $b_1$, and $y_{\mathrm{mid}}$. Each pixel $i$ in the observed profile is weighted by an estimated measurement uncertainty $\sigma_i$. We calculate this uncertainty as $\sigma_i = \sqrt{\sigma_{\mathrm{phot},i}^2 + \sigma_{\mathrm{back}}^2}$, where $\sigma_{\mathrm{phot},i}$ is the photon noise on that pixel and $\sigma_{\mathrm{back}}$ is the background noise at the radial location of the profile due to residual speckles and sky background. The photon noise cannot simply be measured from the observed profile because LOCI self-subtraction has created ``negative-brightness regions'' that distort the photon counts. Instead, we estimate the photon noise as the square root of the pixel's photon count in the model scene $g$ constructed for each least-squares iteration and propagate this uncertainty through the modeling algorithm. This is only a rough estimate that attempts to account for the photon noise per pixel before it was distorted by self-subtraction. The background noise is calculated as the standard deviation of the pixel brightness in the regions of the observed profile that contain neither disk signal nor self-subtraction.

% Model image stack for K.
\begin{figure*}[ht]
\centering
\epsscale{1.0}
\plotone{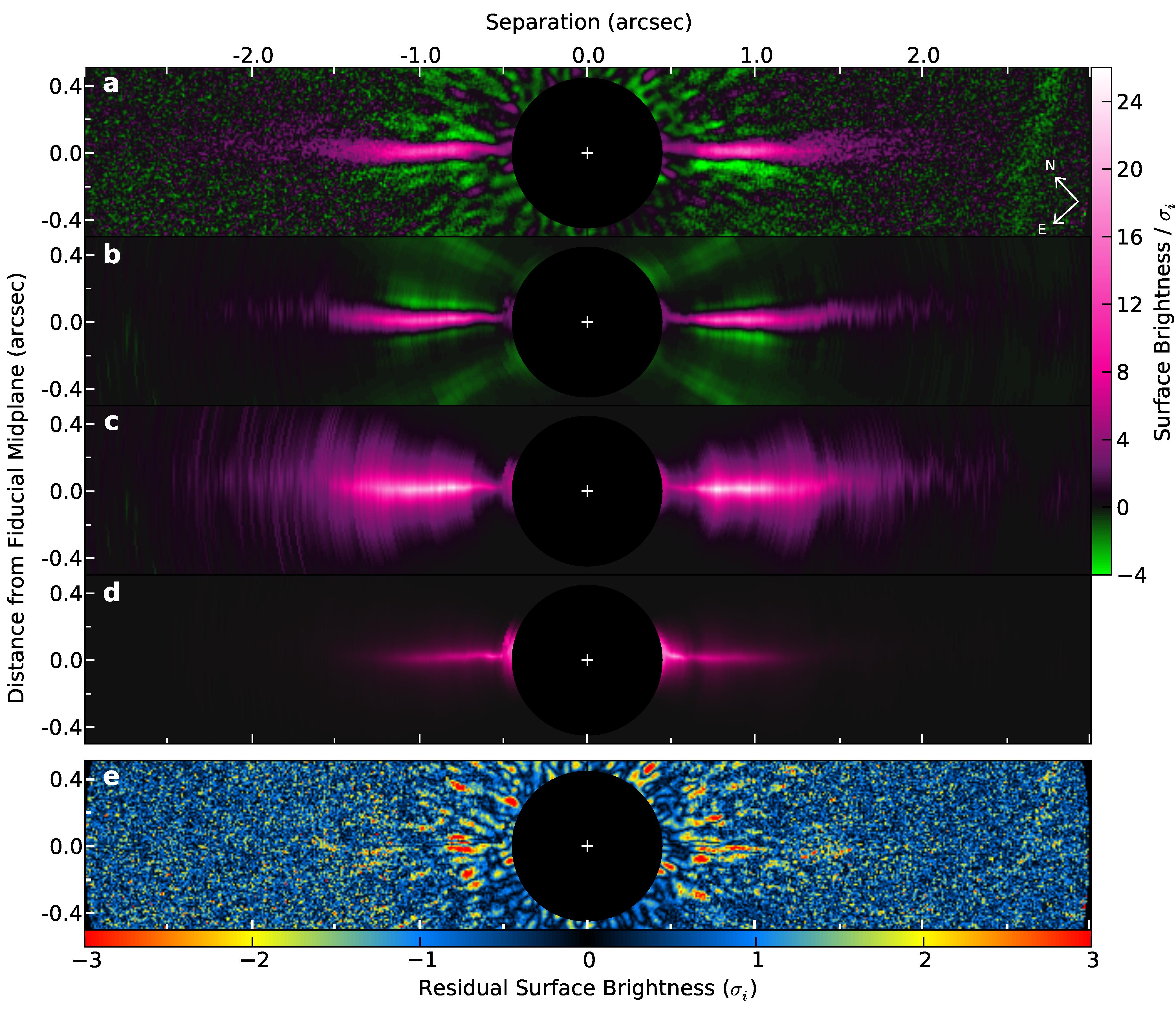}
\caption{$K_s$ images of the disk surface brightness directly analogous to $H$ images in Figure \ref{fig:Hmodels}. The brightness scale for (\textit{d}) is the same as in $H$ but the scale for (\textit{a}) through (\textit{c}) is slightly different than in $H$ to better show the disk features.}
\label{fig:Kmodels}
\end{figure*}

We perform the least-squares fit simultaneously on profiles from LOCI-processed images of varying $N_\delta$ at each separation, which means that a separate model $F$ is constructed for each $N_\delta$. For our $H$ data, we use eight images with $N_\delta$=\{0, 0.5, 1, 1.5, 2, 3, 4, 5\}, where $N_\delta=0$ is the most aggressive reduction (all reference images allowed) and $N_\delta=5$ is the most conservative (few reference images allowed). For our $K_s$ data, we use nine images with $N_\delta$=\{0, 0.25, 0.5, 1, 1.5, 2, 3, 4, 5\}, with the $N_\delta=0.25$ image included because it showed the highest overall S/N for this dataset. All images are weighted equally. This inclusion of data processed with multiple $N_\delta$ values helps the algorithm select a model scene $g$ that accurately fits the data over a wide range of self-subtraction levels. The parameter space of the fit is restricted in two ways. We require that: (1) the midplane position be within the range $y$ of a given profile; and (2) $b_1 \geq b_2$, to maintain consistency within the monotone cubic spline interpolator. The least-squares algorithm produces a covariance matrix $\Sigma$ for the parameters $b_0$, $b_1$, and $y_{\mathrm{mid}}$.

Examples of the best-fit models $F$ and $g$ for a simultaneous fit to vertical profiles at the same separation but from various $N_\delta$ reductions of the $H$ data are shown in Figure \ref{fig:slices}. The underlying scene $g$ is constant for all panels but the self-subtraction function differs for each $N_\delta$, resulting in different $F$ functions.

The best-fit model function $g$ is used to derive the disk brightness and FWHM. We calculate a one-dimensional integrated brightness \citep{graham2007} by integrating $g$ over all $y$. We calculate the disk FWHM directly from $g$ in a straightforward fashion. These parameters are more relevant to the disk's physical properties than the fit parameters $b_0$ and $b_1$. The midplane position is taken directly from the least-squares fit as the best-fit value for $y_{\mathrm{mid}}$.

It is important to note that, because the vertical profile model takes self-subtraction into account, the best-fit values retrieved for the brightness, projected FWHM, and midplane location are those of the model disk \textit{prior to self-subtraction}. Assuming that the model is an accurate description of the disk, then we gain estimates of the true disk parameters, uncorrupted by LOCI processing. The procedure just described in this section for HD 32297 only models the disk at a single radius, so we repeat it for all radii and for both ansae to construct the full two-dimensional brightness distribution, in the process obtaining radial profiles for one-dimensional integrated brightness, projected FWHM, and midplane position.

\subsubsection{Uncertainties in Brightness Modeling} \label{sect:modeling_uncertainty}

We determine the uncertainties on our derived disk parameters from transformations of the covariance matrices ($\Sigma$) associated with the original parameters from the least-squares fits. Correlations exist between the derived parameters, and to transform the variables along with their uncertainties we must account for the Jacobian of the transformation. To do this, we calculate $\Sigma' = J \cdot \Sigma \cdot J^T$, where $\Sigma' $ is the transformed covariance matrix for the derived parameters and $J$ is the Jacobian, which we compute numerically. Finally, taking the square root of the variances along the diagonal of $\Sigma'$ gives the marginal distribution $\sigma$ for all three derived parameters. These uncertainties include contributions from both photon noise and background noise because those terms compose the least-squares weights.

Estimating parameter uncertainties from the least-squares covariance matrices assumes independent, Gaussian-distributed measurement uncertainties. For simplicity in this proof-of-concept application, we will work under this assumption, although we recognize that speckles are not Gaussian-distributed (e.g., \citealt{fitzgerald2006}) and that spatial correlations exist between image pixels. We explore the effects of the non-Gaussian speckle distribution on our uncertainty estimates near the end of this section. Despite these issues, we generally find that derived-parameter measurements in adjacent separation bins agree within their 1$\sigma$ errors (Section \ref{sect:results}). This suggests that our errors properly represent pixel-to-pixel variations in the data, even in the low S/N regions of the disk.

However, we note the important caveat that there may be systematic errors for which our 1$\sigma$ uncertainties do not fully account. One possible source of such systematic error would be an inability of our spline model for the disk vertical profile to match the functional form of the profile. Such a situation could lead the least-squares fit parameters to be over-constrained, consequently causing variances in $\Sigma$ (and therefore in $\Sigma'$) to underestimate the true parameter variances. This should be taken into account when evaluating the significance of features in the radial profiles of Section \ref{sect:results}. Violation of our assumption that the disk brightness is symmetric around the midplane is another potential source of systematic error. An improved model of the disk vertical profile could reduce these errors, which remain preferable to the relatively large uncertainties introduced by self-subtraction and removed by our modeling process.

One way we attempt to mitigate underestimation of systematic errors is by scaling the covariance matrix by the reduced chi-square value ($\chi^2_\nu$). With mean $\chi^2_\nu$ values of $\sim$4.8 for $K_s$ model fits and $\sim$6.0 for $H$ model fits, this scaling increases the associated $\sigma$ errors on our derived disk parameters by factors of $\sim$2.2$-$2.4 on average. In calculating $\chi^2_\nu$, we exclude the residuals from pixels that do not contain either disk signal or self-subtraction. These pixels contain only random noise and are best fit by zero brightness in the model profile. Such ``empty'' pixels make up a significant fraction of each profile and may bias the $\chi^2_\nu$ downward if not excluded, thereby implying a better goodness of fit than is deserved. The disk and self-subtraction signals are spatially correlated due to the diffraction-limited imaging resolution, so we determine the number of degrees of freedom ($\nu$) as the number of resolution elements minus the number of fit parameters. The number of resolution elements is approximated as the number of pixels in the profile divided by the diffraction limit (in pixels). This is a more accurate estimate of the degrees of freedom because our model will only be sensitive to features larger than the diffraction limit.

The $\chi^2_\nu$ metric is itself uncertain, so we must consider how our scaling of the parameter variances by $\chi^2_\nu$ introduces additional uncertainty. The standard deviation for a chi-square distribution is proportional to $\frac{\sqrt{2}}{\nu}$. Conservatively assuming that the quasistatic speckles follow an exponential distribution, we find we need $\nu \gtrsim 200$ for a given semiannulus in order to achieve a fractional uncertainty on $\chi^2_\nu$ of $\lesssim20$\%. Semiannuli only contain this number of degrees of freedom at $r > 290$ AU. Consequently, there is uncertainty in the scaling of our errors, particularly at small separations where speckle noise is prominent. We do not include this additional uncertainty in our error bars but it should be considered when evaluating the significance of our measurements.

We reiterate that our goal in this work is not to produce a perfect representation of the disk, but rather to construct a model that is useful for verification of our self-subtraction modeling method and for estimating the disk's physical parameters. The estimates can later be used as starting points for constructing more complex three-dimensional models that may provide more accurate measurements of those parameters.

% Modeling Results
\subsection{Modeling Results} \label{sect:results}

We present the model surface brightness distributions produced by the process outlined in the previous section for $H$ and $K_s$ bands in Figures \ref{fig:Hmodels} \& \ref{fig:Kmodels}, respectively. For display purposes, panels (a), (b), and (c) have been scaled by $\sigma_i$, which is the separation-dependent random error in the data at each pixel $i$, as calculated in Section \ref{sect:disk_selfsubmodel}. Note that this differs from $\sigma$ described in Section \ref{sect:modeling_uncertainty}. Panel (a) shows the LOCI image from Figure \ref{fig:LOCI_finals} with a different color stretch and rotated counterclockwise by $42.5^\circ$ so that the midplane lies along the horizontal. In panel (b) is a model with self-subtraction included, constructed by solving Equation \eqref{eq:F4} for the best-fit parameter values at all radii. Negative-brightness regions created by self-subtraction are visible above and below the midplane in the model image, similar to the LOCI-processed data in location and amplitude. Panel (c) shows a model of the disk as it appeared before corruption by self-subtraction and random error, which is constructed from $g(r, \theta)$. The banded structures of the models are artifacts primarily due to fluctuations in the best-fit parameters of the one-dimensional models at neighboring separations. Panel (d) shows the same model as in (c) but with no scaling. Finally, (e) is a deviate map calculated as [(a) - (b)]/$\sigma_i$. The deviates were not scaled by $\chi^2_\nu$ and thus do not account for correlated pixels or over-constrained model parameters. We note that the deviates are $>$$3\sigma_i$ in some parts of the disk, indicating imperfect agreement between data and model (see Section \ref{sect:fidelity}).

By inspection, the pre-self-subtraction models (c) are consistently brighter than the self-subtracted models (b) along the midplane and in the wings. This indicates that we have recovered disk brightness that was previously lost to self-subtraction. We use the pre-self-subtraction model to derive radial profiles for the disk brightness, projected FWHM, and midplane location.

% Brightness profiles results
\subsubsection{Brightness Profiles} \label{sect:brightness}

% Model brightness profiles for H and K.
\begin{figure}[ht]
\centering
\epsscale{1.2}
\plotone{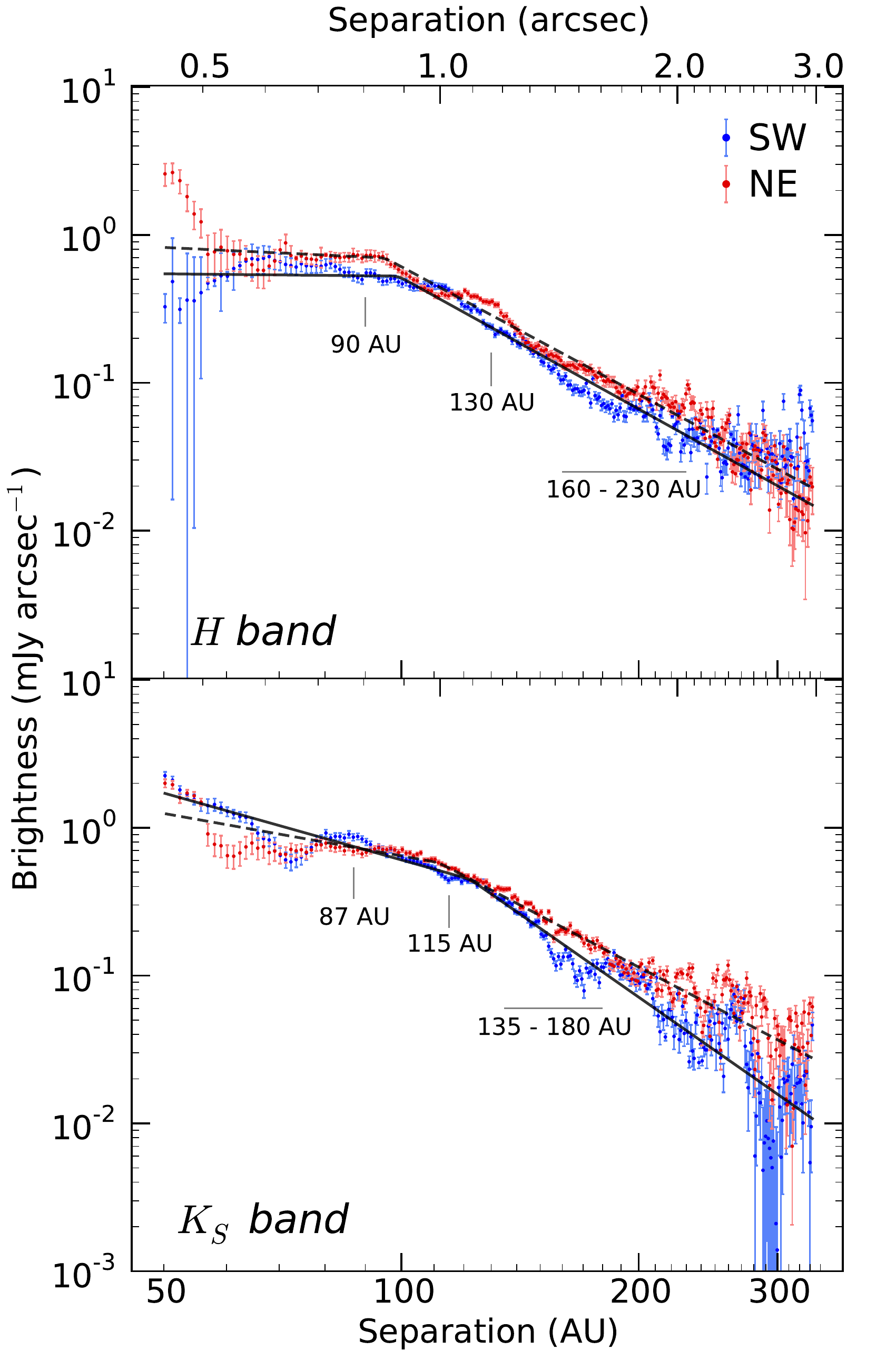}
\caption{One-dimensional integrated brightness profiles produced by our self-subtraction modeling process in $H$ and $K_s$. Measurements represent the disk brightness before flux loss caused by ADI/LOCI processing. The SW ansa is in blue and the NE is in red. Locations of significant asymmetries are labeled. The error bars represent 1$\sigma$ uncertainty levels that include $\chi^2_\nu$ scaling and uncertainties associated with our simplified disk model, with the caveat that the errors may remain underestimated due to over-constrained least-squares fit parameters. The best-fit power-law functions are shown as solid lines for the SW and dashed lines for the NE. Breaks in the power laws occur at $r\approx95$$-$125 AU. Low S/N in the $K_s$ data reduces the reliability of measurements at $r\gtrsim160$ AU.}
\label{fig:modelBrights}
\end{figure}

Figure \ref{fig:modelBrights} shows the radial brightness profiles produced by our modeling of the final LOCI-processed images in both bands. We plot the one-dimensional integrated brightness (mJy arcsec$^{-1}$), which is intrinsically absent of self-subtraction because the modeling process accounts for radius-dependent flux loss. The 1$\sigma$ errors on the measurements were derived from covariance matrices scaled by $\chi^2_\nu$, thereby accounting for some of the systematic errors related to our simplified disk model (see Section \ref{sect:modeling_uncertainty} for details). This applies to our FWHM, midplane location, and color measurement errors in the following sections, as well.

The profiles extend inward to a separation of 50 AU ($0.45''$) in both bands. Interior to this point the disk signal is dominated by residuals from incomplete PSF subtraction, and uncertainties on the brightness in this region are large. The profiles continue out to $\sim$335 AU ($3.0''$) in $H$ and $\sim$160 AU ($1.4''$) in $K_s$. Beyond these separations the disk brightness decreases to the level of the background noise and the disk is not detected. The S/N of the $K_s$ data is lower than that of the $H$ data at $r\gtrsim160$ AU, leading to lower confidence in the $K_s$ profiles at these larger separations.

We see brightness features in both ansae, but, as is inherent in all high-contrast disk imaging, comparison to multiple independent data sources is necessary because each telescope, instrument, observational technique, and data reduction method has its own distinct systematic errors. There is an additional complication that different groups use different techniques for measuring brightness (e.g., flux in apertures of varying size, or midplane brightness). We note that all of the previous works discussed below used apertures of various sizes to extract surface brightness profiles, so our profiles are analogous but not directly relatable to theirs in an absolute sense (see Section \ref{sect:apertures} for further discussion).

The $H$-band brightness features are summarized here according to separation:

\noindent $\mathbf{r \lesssim 55}$ \textbf{AU:} We measure an abrupt rise in NE brightness toward smaller radii. This could be due to contamination by speckle noise that is not adequately captured by our error bars (see Section \ref{sect:modeling_uncertainty} for details) or it could be similar to the steep increases reported by \citet{schneider2005} and \citet{currie2012_32297} at 1.1 $\mu$m and in $K_s$, respectively. However, we find the NE to be brighter than the SW in this region, which is opposite the asymmetry noted by the other two works.

\noindent $\mathbf{r\approx}$ \textbf{90 AU, 130 AU, 160$-$230 AU:} We find a NE $>$ SW asymmetry at these locations. The asymmetry at 130 AU coincides with a LOCI artifact on the NE ansa (possibly caused by a sharp transition between adjacent subtraction subsections with different noise characteristics) and may be an artificially-produced increase in the NE brightness. \citet{boccaletti2012} presented $H$ profiles that showed no statistically significant asymmetries. They processed their ground-based data with a ``classical ADI'' that they considered less aggressive than LOCI in order to try to preserve photometric fidelity. Our findings differ from the asymmetry detected at 1.6 $\mu$m by \citet{debes2009}, who noted the SW ansa to be significantly brighter than the NE ansa at $r\approx112-280$ AU. Other space-based measurements by \citet{schneider2005} at the shorter wavelength of 1.1 $\mu$m showed no statistically significant asymmetries between 56 AU and 190 AU but those authors did report the SW ansa to be systematically brighter than the NE at $r\approx157-235$ AU.

In $K_s$, we find the ansae to be generally symmetric, which agrees with \citet{boccaletti2012}, who presented ground-based $K_s$ profiles that were symmetric between 56 AU and 336 AU. However, we do find asymmetries at these locations:

\noindent $\mathbf{r\approx60}$ \textbf{AU, 87 AU:} We see SW $>$ NE asymmetries at these separations, though the former feature is located at the edge of the speckle-dominated region.

\noindent $\mathbf{r\approx115}$\textbf{ AU:} We see a marginally significant NE $>$ SW asymmetry here.

\noindent $\mathbf{r\approx}$ \textbf{135$-$180 AU:} We note a NE $>$ SW asymmetry in this low confidence region. Partially overlapping this separation at $r\approx120-145$ AU, profiles from \citet{currie2012_32297} indicated a marginally significant NE $>$ SW asymmetry in $K_s$, which was not remarked on. In contrast, \citet{debes2009} noted a significant SW $>$ NE asymmetry at 2.05 $\mu$m in the region $r\approx112-280$ AU.

\noindent Additionally, although we consider the $r > 275$ AU region insignificant for this band, we note that the dip in $K_s$ SW brightness there is not a failure of the fitting algorithm but suppression of the disk in the LOCI image by an artifact present in several images of the dataset.

On average in the highest confidence region of $r\approx60-160$ AU, the SW ansa is $\sim$36\% brighter in $K_s$ than in $H$ and the NE is $\sim$23\% brighter in $K_s$. This may be a real feature born of the disk's dust distribution or, as discussed in Sections \ref{sect:fidelity} and \ref{sect:fwhm_discussion}, it could be due to extended wings in the model vertical profiles that increase the model FWHM and thereby increase the integrated brightness.

In general, we find the brightness profiles to be largely symmetric between ansae, but there are several regions of NE $>$ SW asymmetry. Previous ground-based observations show similar asymmetries \citep{currie2012_32297} and elongation of the NE ansa \citep{currie2012_32297}. These results differ from those of the NICMOS observations by \citet{schneider2005} and \citet{debes2009}, which generally show the SW to be brighter than the NE. Such discrepancies between ground-based and space-based measurements underscore the different systematic uncertainties associated with both types of imaging. We have already discussed the error sources associated with AO observations processed with ADI/LOCI. Although the NICMOS data do not suffer from ADI self-subtraction, they do not necessarily present unbiased photometry because they could contain other image-processing artifacts from a variety of sources (e.g., spectral mismatch between PSF reference star and HD 32297, telescope's ``breathing'' \citep{fraquelli2004}, unsubtracted thermal background, registration errors between stacked images). Systematic errors could impact the NICMOS photometry, and therefore affect asymmetries in radial brightness profiles or disk colors (see Section \ref{sect:color}).

% Brightness profile power-law results
\subsubsection{Brightness Profile Power-Law Fits} \label{sect:brightness_powerlaws}

To characterize the change in brightness as a function of radius, we fit broken power laws to each profile and tabulated the best-fit results in Table 1. We fit to the entire profile in each case. The relevant parameters are the location of the break ($r_{\mathrm{break}}$), power-law index interior to the break ($\alpha$), and power-law index exterior to the break ($\beta$). The uncertainties shown represent the 1$\sigma$ level. In $H$, the measurements of all three parameters are consistent between the two ansae. In $K_s$, $r_{\mathrm{break}}$ occurs $\sim$10 AU farther out in the SW than the NE. The indices $\alpha$ and $\beta$ are also significantly steeper for the SW than the NE. In comparing the two ansae by wavelength, we find the SW and NE breaks to occur $10-20$ AU farther from the star in $K_s$ than in $H$. The $K_s$ profiles also show steeper $\alpha$ values and a steeper SW $\beta$ value than the $H$ profiles. 

% Adjust vertical spacing around table to reduce white space.
\vspace{-5mm}
% Model brightness power-law best-fit parameter table.
\setcounter{table}{1}
\begin{center}
\begin{table}[ht]
\begin{tabular}{l c c c c}
\multicolumn{5}{c}{TABLE 1} \\
\multicolumn{5}{c}{Model Brightness Profile Power Law Best-fit Parameters} \\
\hline
\hline
Band & Ansa & $r_{\mathrm{break}}$ (AU) & Inner Index ($\alpha$) & Outer Index ($\beta$)  \\
\hline
\multirow{2}{*}{$H$} & SW & $99.1\pm2.0$ & $-0.05\pm0.20$ & $-2.95\pm0.05$ \\
 & NE & $95.3\pm2.1$ & $-0.25\pm0.28$ & $-2.87\pm0.03$ \\
 \hline
\multirow{2}{*}{$K_s$} & SW & $122.2\pm1.9$ & $-1.50\pm0.07$ & $-3.73\pm0.07$ \\
 & NE & $111.4\pm1.9$ & $-0.95\pm0.10$ & $-2.79\pm0.04$ \\
\hline
\label{tab:PLfits}
\end{tabular}
\end{table}
\end{center}
\vspace{-5mm}

We can compare our power-law results with those of previous works, again with the caveat that our profiles were derived in a different manner from the profiles in those works. Our $H$-band results for $r_{\mathrm{break}}$ are roughly consistent with those presented by \citet{debes2009} for NICMOS 1.6 $\mu$m measurements, though our inner and outer indices are consistently shallower than theirs. \citet{boccaletti2012} found $r_{\mathrm{break}}$ and $\beta$ for their ``classical ADI'' $H$ data that are approximately consistent with our own. Their $\alpha$ are positive and substantially different from our values, though the authors note that these measurements are uncertain due to low S/N in the inner regions of their $H$ data. Perhaps due to differences in wavelength or instrument, our measurements differ from the shorter 1.1 $\mu$m results in \citet{schneider2005}. That work found a single power law in the SW, though the index (-3.57) was very similar to our outer index. They fit a broken power law to the NE that had a greater $r_{\mathrm{break}}$ (190 AU) and a steeper inner index (-3.7) than ours, but found a similar outer index (-2.74). At still shorter wavelengths, $R$-band profiles from \citet{kalas2005_32297} yielded single power law indices of $-3.1\pm0.2$ and $-2.7\pm0.2$ for the SW and NE ansae, respectively, at $r\approx5''-15''$, which, if extrapolated inward, would be similar to our outer indices.

With respect to the $K_s$-band results, our $r_{\mathrm{break}}$ values roughly agree with 2.1 $\mu$m values from \citet{debes2009}. Our SW $\beta$ and NE $\alpha$ are also consistent with their findings, but our SW $\alpha$ is steeper by a factor of $\sim$3 and our NE $\beta$ is $\sim$1.4 times shallower than theirs. The \citet{boccaletti2012} $K_s$ SW parameters are similar to the \citet{debes2009} SW values, thus our SW results compare similarly. In the NE, we record a similar $r_{\mathrm{break}}$ and $\alpha$, but a shallower $\beta$ (by a factor of $\sim$1.6) than reported by \citet{boccaletti2012}. Finally, \citet{currie2012_32297} found they could not fit a power law to their $K_s$ profiles inward of $\sim$112 AU, which is roughly consistent with the location of our break. However, they found much steeper power law indices ($< -5.1$) than ours exterior to that point, whether it be a single power law (for the SW) or a broken power law (for the NE).

\subsubsection{Disk Width} \label{sect:fwhm}

\begin{figure}[ht]
\centering
\epsscale{1.2}
\plotone{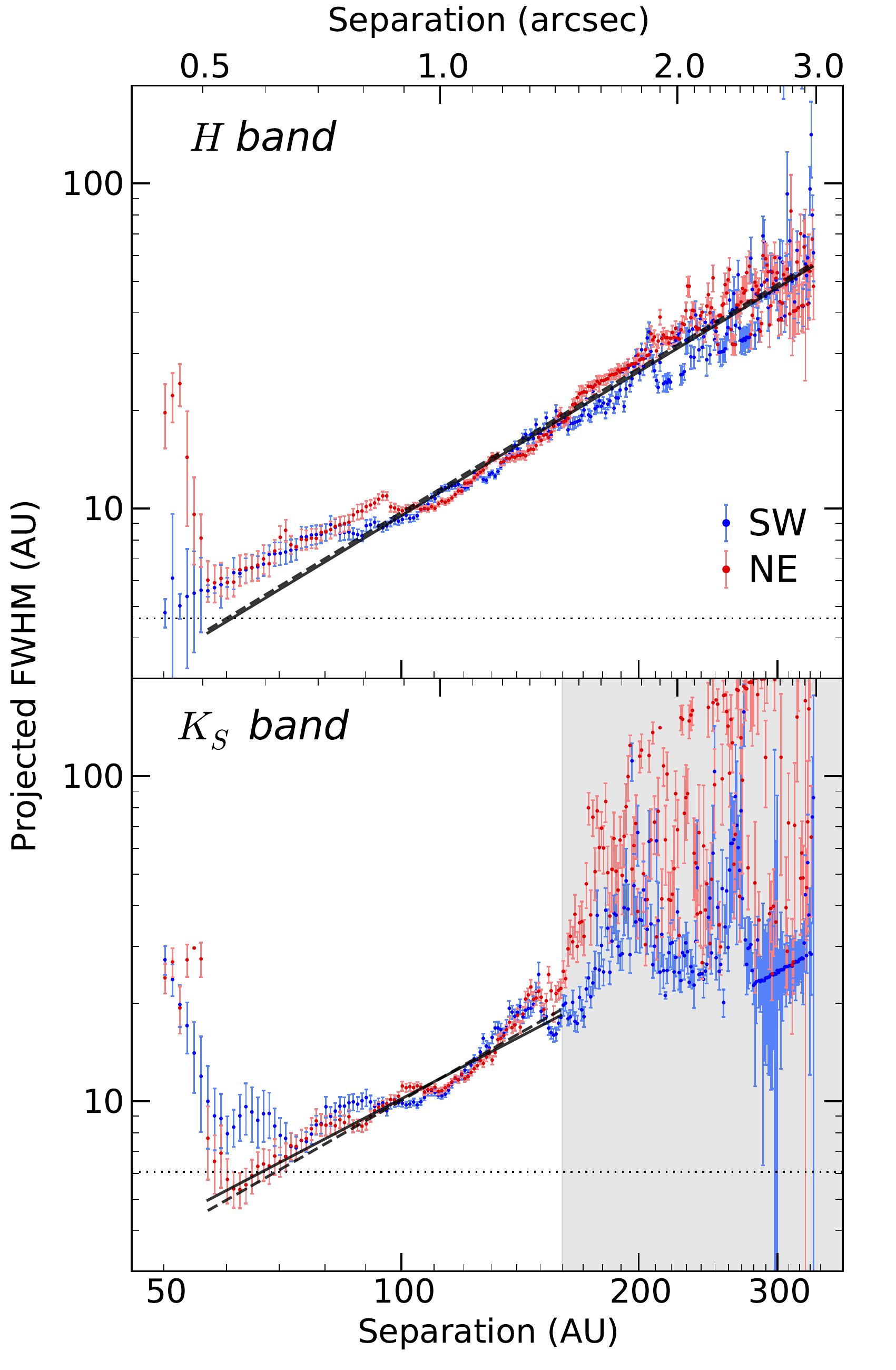}
\caption{Projected FWHM of the pre-self-subtraction model disk as a function of separation for SW and NE ansae in $H$ and $K_s$. We find the FWHM to generally increase with separation. The best-fit power-law functions are shown as solid lines for the SW and dashed lines for the NE. The gray-shaded area in $K_s$ marks the low confidence region that we exclude from most of our results and discussion. The horizontal, dotted line indicates the diffraction limit (1.22$\lambda$/D) at each band's central wavelength. The error bars represent 1$\sigma$ uncertainty levels that include $\chi^2_\nu$ scaling and uncertainties associated with our simplified disk model.}
\label{fig:modelFWHM}
\end{figure}

As part of the self-subtraction modeling process, we also extracted radial profiles for the disk projected FWHM, where the FWHM is measured from the pre-self-subtraction vertical profile model ($g$) at a given radial separation. We plot these radial profiles in Figure \ref{fig:modelFWHM}. In both bands, the disk FWHM consistently increases with separation. The only region in which this doesn't hold is at $r\lesssim60$ AU, where speckles contaminate the disk and make it appear unusually broad. \citet{currie2012_32297} also found FWHM to increase with separation. \citet{boccaletti2012} noted that the FWHM of the $K_s$ ADI/LOCI-processed disk increased with separation and determined, as we do, that model disks that were not affected by self-subtraction had generally larger FWHM than the processed data.

% Model FWHM profiles for H and K.

In $H$, the FWHM ranges from 5 AU to $\sim$60 AU. In $K_s$, the FWHM ranges from 6 AU to $\sim$22 AU (in the region $r=60-160$ AU). At $r>160$ AU, the low S/N in the $K_s$ image severely affected the model-fitting algorithm, leading to disk models that were unbelievably wide. That the error bars do not fully account for the scatter in this region is possible evidence that the errors remain underestimated due to systematics involved with our model choice. It could also indicate a violation of our Gaussian distribution assumption. Qualitatively, we find that our $K_s$ FWHM values are generally smaller than those of \citet{currie2012_32297} and greater than the observed LOCI-reduced FWHM measurements of \citet{boccaletti2012} at a given separation. We find no highly significant differences in FWHM between ansae in either band, similar to \citet{currie2012_32297}. Perhaps serendipitously, we see a bump in the $K_s$ SW profile at $r\approx67$ AU that corresponds to a similar bump in \citet{currie2012_32297}, though again the significance is low.

However, we do find the FWHM to be an average of seven percent wider in $K_s$ than in $H$ at $r\approx60$$-$160 AU. This may be a true difference in the disk's appearance in the two bands or it may be a result of our modeling, as we discuss in Sections \ref{sect:fidelity} and \ref{sect:fwhm_discussion}. The drop to FWHM $\approx$ 38 AU in the $K_s$ SW profile at $r>275$ AU is due to the same artifact discussed in the previous subsection. Finally, we note that the FWHM is greater than or statistically consistent with the resolution threshold set by the diffraction limit at all separations.

We fit single power laws to the FWHM profiles to further investigate the increase in disk width as a function of separation. For both bands we fit the profiles to a minimum separation of 57 AU because inward of there the FWHM is likely biased by speckles. We fit the $H$ profile to its outermost separation and truncated the $K_s$ profile at $r=160$ AU for fitting purposes to avoid the low confidence region. We find best-fit power-law indices of 1.47 $\pm$ 0.04 for both ansae in $H$, 1.27 $\pm$ 0.14 for $K_s$ SW, and 1.38 $\pm$ 0.11 for $K_s$ NE.

\subsubsection{Midplane Position} \label{sect:midplane}

\begin{figure}[ht]
\centering
\epsscale{1.2}
\plotone{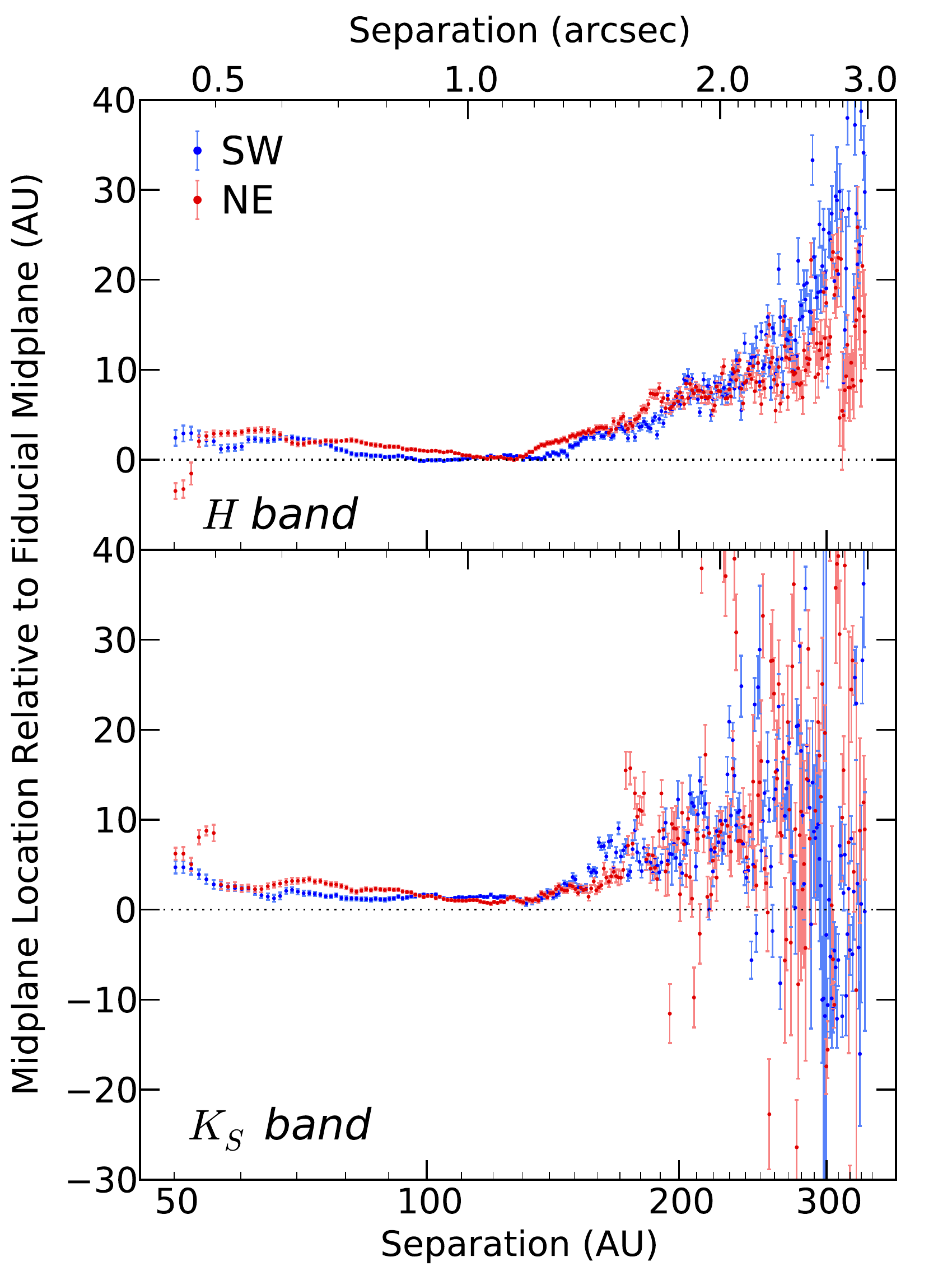}
\caption{Disk midplane location measured by the self-subtraction modeling algorithm and relative to the fiducial midplane location (PA=$47.5^\circ$ east of north; dotted line) as a function of separation for SW and NE ansae in $H$ and $K_s$. A positive location corresponds to a position northwest of the fiducial midplane as it extends from the star and a negative location is southeast of the fiducial midplane. The profiles show considerable northwest curvature of the disk, particularly in $H$. The error bars represent 1$\sigma$ uncertainty levels that include $\chi^2_\nu$ scaling and uncertainties associated with our simplified disk model.}
\label{fig:modelMidplane}
\end{figure}

Figure \ref{fig:modelMidplane} shows radial profiles for the disk midplane position relative to a fiducial midplane with PA $=47.5^\circ$ east of north. We locate this fiducial midplane based on visual inspection of our LOCI images and previous PA measurements \citep{debes2009, boccaletti2012}. We find that the midplane position moves farther toward the northwest as $r$ increases, indicating bowing or curvature of the disk in that direction. This curvature begins to become more pronounced at $r\approx130$ AU and is approximately equal in degree in both bands, with a maximum deviation of $\sim$30 AU at the largest separations, although there is considerably more scatter in $K_s$ at large separations where S/N is low. We also find that the midplane most closely approaches the fiducial midplane at intermediate separations of $r\approx80$$-$130 AU. Interior to this region, the midplane is again generally located farther to the northwest than the fiducial position.

Our results agree with previous reports of curvature in the disk. \cite{boccaletti2012} reported a midplane deviation of a few AU to the north, particularly on the NE ansa in $K_s$, at $r\approx67-112$ AU. This agrees approximately in amplitude and location with the non-zero deviation that we find for the midplane position of the inner disk. Northward curvature was also noted by \citet{currie2012_32297} at separations of $r\geq 100$ AU that is similar to the curvature we find in the outer disk. \citet{debes2009} reported westward curvature in the SW ansa (and suggested similar curvature in the NE ansa) at separations of 0.5$''$$-$3.0$''$, something that \citet{boccaletti2012} also possibly detected. Previously, \citet{kalas2005_32297} had noted that the SW disk emission curved west with radius at separations greater than 5.0$''$. We discuss possible causes of the curvature in Section \ref{sect:disc_curv}.

% Model midplane location profiles for H and K.

% Model color profiles for H-K.
\begin{figure}[ht]
\centering
\epsscale{1.2}
\plotone{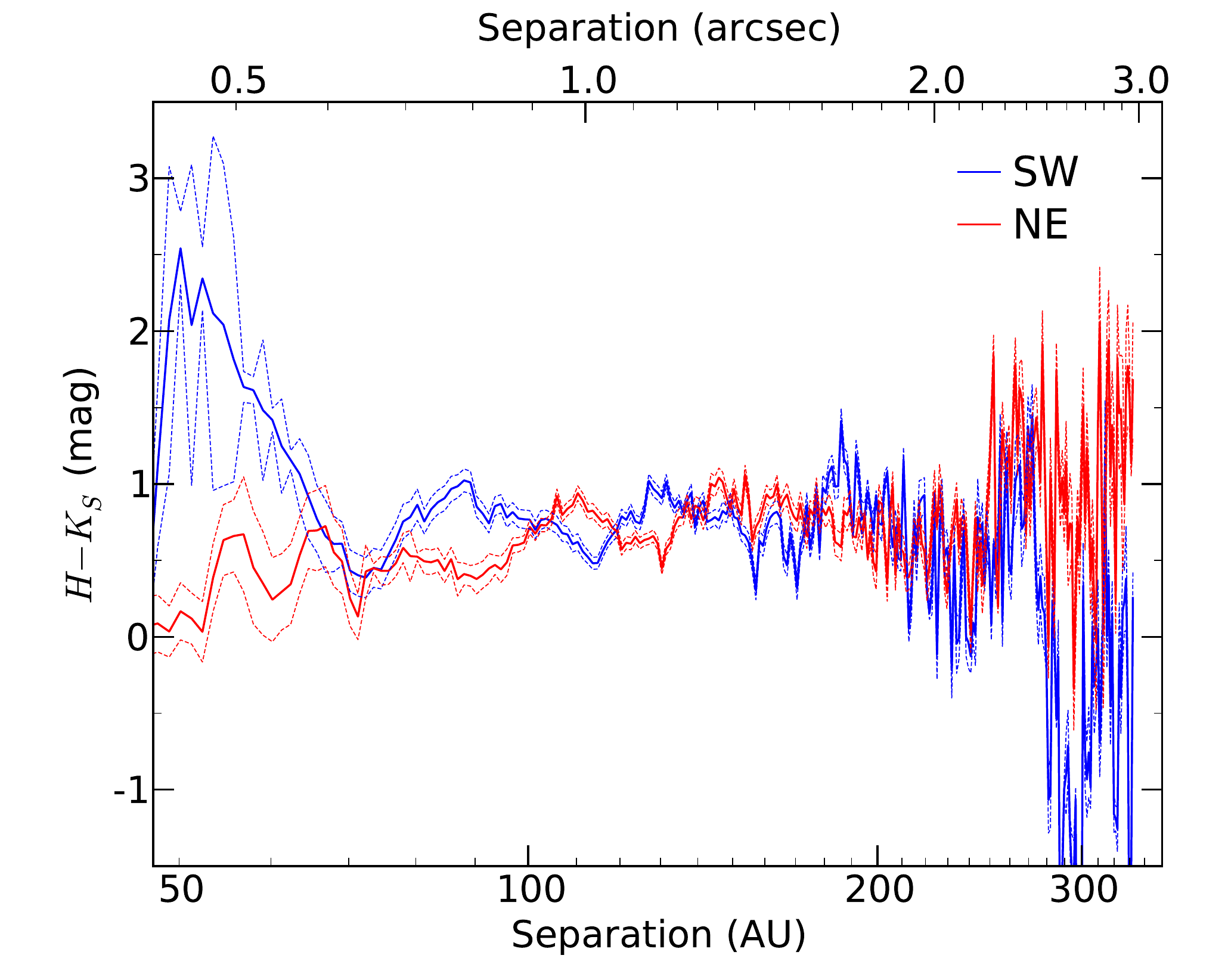}
\caption{Disk $H-K_s$ color (solid lines) relative to the star as a function of radius, computed from self-subtraction-corrected one-dimensional brightness measurements. The SW ansa is in blue and the NE is in red. Dashed lines show the 1$\sigma$ confidence intervals, which include $\chi^2_\nu$ scaling and uncertainties associated with our simplified disk model. The disk is generally red and its color is approximately constant with $r$. We have lower confidence in our measurements at $r\gtrsim160$ AU because of the low S/N in $K_s$ in that region.}
\label{fig:modelColor}
\end{figure}

\subsubsection{Disk Color} \label{sect:color}

Using the one-dimensional brightness profiles presented above, we computed the $H-K_s$ color of the disk, which is shown in Figure \ref{fig:modelColor}. We normalized the disk color by subtracting the corresponding stellar color ($H=7.62$, $K=7.59$; Two Micron All Sky Survey; \citealt{cutri2003}). The disk is generally red, with 0 mag $\lesssim H-K_s \lesssim$ 1 mag at all separations. In the highest confidence region of $r<160$ AU, we do not detect any highly significant color difference between the two ansae, nor do we find the disk to be particularly red at any single separation.

The red color of the disk in our data is most consistent with results from \citet{boccaletti2012} when compared with other works. \citet{boccaletti2012} mentioned that the disk was dimmer in $H$ than in $K_s$ and their surface brightness profiles (their Figure 6) showed $K_s > H$ throughout much of the disk, though often with low statistical significance. Our results are less consistent with the conclusions of \citet{debes2009}, who inferred colors from the dust's normalized scattering efficiency and found the disk to be generally gray in the inner regions and blue at the outer edges over a wavelength range of 1.1$-$2.1 $\mu$m, with the exception of a red zone along the SW ansa between 127 AU and 313 AU. As we noted in Section \ref{sect:brightness}, disagreement of our photometry results with those of \citet{debes2009} could stem from multiple sources of uncertainty. At shorter wavelengths than we investigated, \citet{kalas2005_32297} combined an extrapolated $R$-band profile (based on surface brightness measurements at $r>5''$) with the 1.1 $\mu$m profiles to derive a blue color. Only considering total flux per ansa, \citet{mawet2009} stated that the 1.1 $\mu\mathrm{m}/K_s$ fractional flux ratio (with 1.1 $\mu$m data from \citealt{schneider2005}) was perhaps slightly blue, though statistically consistent with a gray color.

We made an alternative measurement of the disk color using the disk's peak brightness only (which is simply the best-fit value in mJy arcsec$^{-2}$ of the model parameter $b_0$). By this method, we find a slightly blue color at most separations and gray at the rest, which agrees better with the \citet{debes2009} colors than our integrated-brightness results. See Section \ref{sect:color_discussion} for further discussion of the differences between our two color measurements.

As comparing color measurements between different datasets and analysis procedures can be difficult, we leave a more detailed comparison to future work.

\section{DISCUSSION} \label{sect:discussion}

\subsection{Fidelity of Brightness Modeling} \label{sect:fidelity}

In applying our technique to these data, we are modeling two things: (1) the underlying brightness distribution of the disk, and (2) the self-subtraction function. If we first show that we are modeling self-subtraction properly, then we can characterize the accuracy of our recovered brightness distributions.

We tested the ability of our algorithm to model self-subtraction on synthetic ADI datasets in which we knew the underlying disk brightness distribution exactly. We created synthetic brightness distributions based on various vertical profile forms including Gaussian, Lorentzian, and cubic spline functions. These underlying distributions then had one of three levels of noise added: no noise, Gaussian-distributed noise added to each pixel, or the actual stellar PSF and speckle pattern from the $K_s$ dataset. Each dataset was processed in the same manner as the real data using LOCI with a range of $N_\delta$ values and then our self-subtraction modeling algorithm was applied.

As expected, for the cubic spline-based distribution we generally achieved a near-perfect recovery of the self-subtracted final image and also the original underlying brightness distribution, even for the dataset with real speckle noise included. Our radial profiles for integrated brightness, FWHM, and midplane location generally agreed with the synthetic input disk's pre-self-subtraction values to within the 1$\sigma$ errors we calculated. This reproduction of the self-subtraction is by construction, but it serves as a verification of our implementation. We also recovered the Gaussian- and Lorentzian-based distributions but with less accuracy, as our spline model generally underestimated the disk FWHM and overestimated the peak and integrated brightnesses by more than 1$\sigma$. We found that this was largely due to the inability of the simple spline function to perfectly reproduce the forms of the Gaussian and Lorentzian functions. This is discussed more below. Overall, these tests demonstrated that we could accurately forward-model the location and amplitude of self-subtraction in a LOCI-processed image, provided we have an accurate model of the pre-self-subtraction brightness distribution and that the scene is constant across all images. Therefore, we conclude that we are modeling the self-subtraction function correctly.

That leaves us to characterize our ability to model the underlying brightness distribution of our observations. We find that our simple model does not have a sufficient number of parameters to capture all of the observed disk structure. This is most clearly visible in the $K_s$ deviates shown in Figure \ref{fig:Kmodels}e, where there are deviates $>3\sigma_i$ at the location of the disk. We expect that the discrepancies between observations and the model brightness distribution indicate that the best-fit model vertical profile was inaccurate at some separations. This could be partly due to noise or PSF residuals in the data, which contaminated the disk signal and altered its apparent structure. It is likely that the form we chose for the vertical profile model was too simple and was incapable of reproducing the disk shape perfectly, as was the case with our Gaussian and Lorentzian synthetic disks. Recall that the shape of the profile was entirely determined by just two variables ($b_0$ and $b_1$). With a more complex disk model, we could produce a more precise reconstruction of the disk's brightness distribution. The model form is also a potential source of the relatively bright emission far from the midplane in the pre-self-subtraction models ($K_s$ especially). With only one parameter controlling the brightness of the profile's wings, that brightness could be overestimated because much of the information about such extended emission is lost to self-subtraction. Additionally, as our fake-disk tests indicated, our quoted errors may not accurately reflect the uncertainty in the radial profile measurements if the spline model cannot adequately reproduce the true disk shape. Possible solutions to these problems include adding more control points to the spline model or changing to a different functional form (i.e., not a monotonic spline) for the profile.

By applying our forward-modeling technique we are able to reduce the error in recovering the disk brightness to levels lower than the scale previously set by errors from self-subtraction. Therefore, our ability to accurately recover the disk brightness distribution is improved after applying our technique. This is illustrated by a comparison between brightness measurements made via our modeling with those made via aperture photometry in Section \ref{sect:apertures}. We are also able to make more sensible comparisons between different sets of observations, such as our $H$ and $K_s$ datasets. These two datasets included different numbers of images and different amounts of angular rotation, making the effect of self-subtraction differ between them as well. Our self-subtraction modeling puts the final $H$ and $K_s$ images on more equal footing for comparison than they would be otherwise.

The simple modeling that we have done in this work serves primarily as a guide for the construction of more detailed models. Using the rough estimates of the disk parameters that we have derived here, we can construct a three-dimensional model of the dust distribution (left to a future work). From that, we can extract a two-dimensional scattered light distribution, quickly compute the self-subtraction function for that distribution, and combine the two to form a model that can be compared to the LOCI image. We emphasize that computing the self-subtraction function does not require processing the new scattered-light model with LOCI. This saves on computation time when compared to the disk-injection and pre-reduction disk-subtraction methods mentioned in Section \ref{sect:selfsubmodel}, which require LOCI reductions for each new model. We also reiterate that model-data comparisons in the disk-injection method incorporate twice as much speckle noise as the same comparisons in our method because the injected model disk is inserted into the observed data, whereas our model disk can be free of noise and still have self-subtraction applied to it.

\subsection{Degeneracies in Brightness Modeling of ADI Data} \label{sect:robustness}

We have seen that ADI and LOCI filter image data, as discussed in Section \ref{sect:intro}, with the degree of filtering dependent on the amount of field rotation in the dataset and the aggressiveness of the PSF subtraction (which largely depends on the $N_\delta$ LOCI parameter). This filtering means we have incomplete information from which we are trying to recover the disk brightness distribution. Consequently, recovery of that distribution's parameters from a single processed image will be degenerate. Such dependence on image processing methods is common among high-contrast imaging studies. However, we can reduce the degeneracy on some spatial scales by simultaneously fitting image data processed with different values of $N_\delta$. This does require multiple LOCI reductions of the data, but we found just eight and nine reductions to be sufficient for $H$ and $K_s$, respectively (Section \ref{sect:disk_selfsubmodel}). Plus, it is easy to compute the self-subtraction function for different reductions, as only the $N_\delta$ parameter and LOCI coefficients change.

Fitting multiple reductions will not totally remove degeneracy, as the disk emission on the largest azimuthal scales will still be filtered out through ADI image processing. Nevertheless, this approach will help break degeneracy on the scales governed by $N_\delta$.

\subsection{Comparison to Aperture Photometry} \label{sect:apertures}

The primary advantage of our modeled profiles is that they are corrected for self-subtraction, while aperture photometry makes measurements that are more strongly affected by self-subtraction.

% Aperture brightness profiles as function of radius and aperture size for H only.
\begin{figure*}[ht]
\centering
\epsscale{1.15}
\plotone{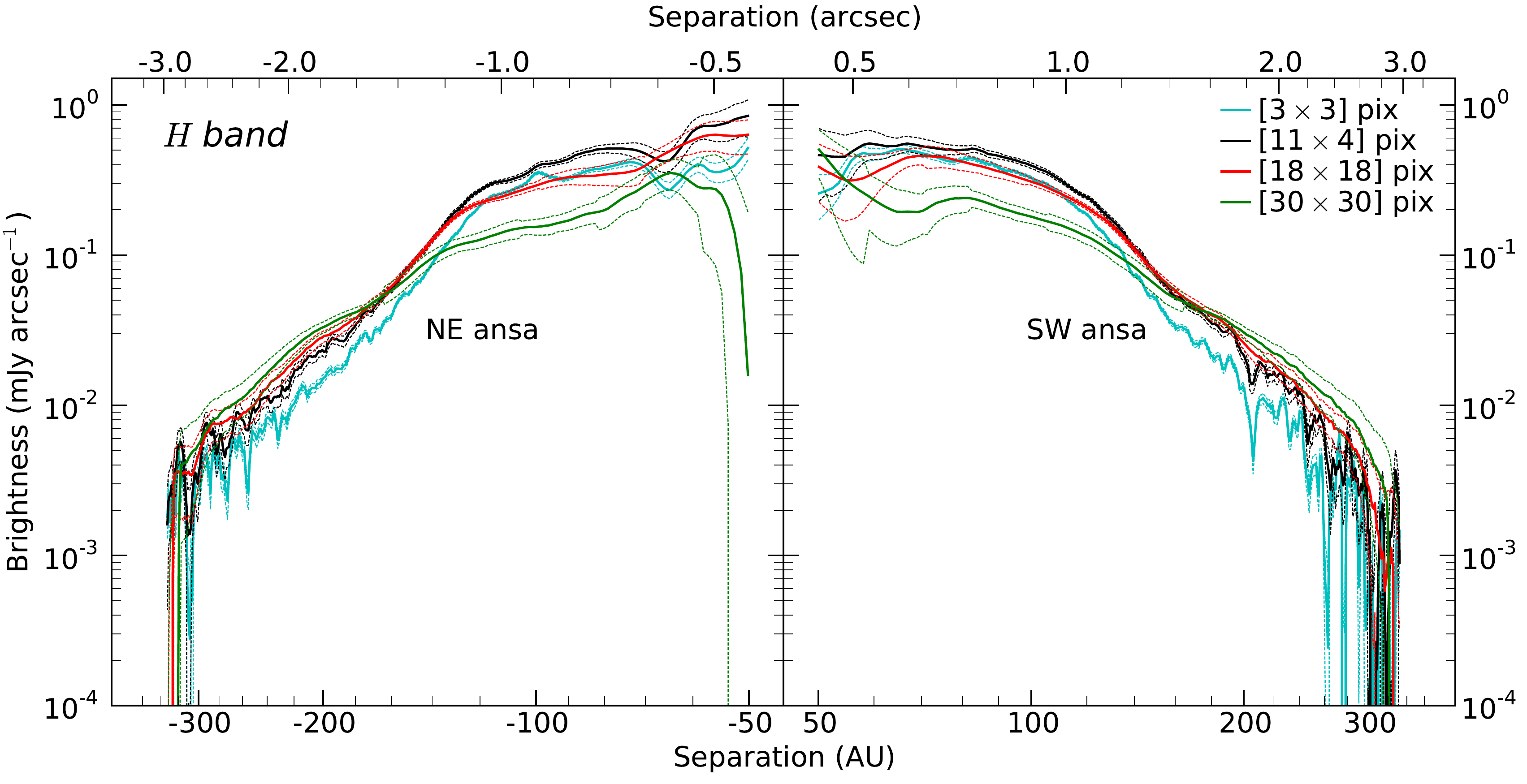}
\caption{Aperture photometry brightness measurements converted to one-dimensional integrated brightness profiles (solid lines) for the $H$-band LOCI final image. Dashed lines show the 1$\sigma$ confidence intervals. Four different aperture sizes (azimuthal $\times$ radial diameter) were used: [3$\times$3] (cyan), [11$\times$4] (black), [18$\times$18] (red), and  [30$\times$30] pixels (green). Brightness values for a given separation can vary between different-sized apertures by factors up to $\sim$2.5. The star is located at $x=0$ (not plotted).}
\label{fig:apBrights}
\end{figure*}

Although aperture measurements are simpler to execute than our self-subtraction modeling method, the choice of aperture size has a large effect on results for an extended source. This is because self-subtraction varies according to both radial and azimuthal position in an image and different-sized apertures will probe different regions of the image. For example, a large aperture that extends beyond the disk and into the self-subtraction-generated negative-brightness regions will capture all of the disk light but also some of the negative brightness, thereby biasing the surface brightness measurements downward. A smaller aperture would yield different results. Additionally, aperture photometry is sensitive to variations in disk width and the degree of self-subtraction with projected separation. In contrast, the modeled profiles are relatively immune to these factors. We attempted to quantify the effect of aperture size on the surface brightness profile for our highest S/N $H$-band image in Figure \ref{fig:apBrights}.

We calculated the mean surface brightness in rectangular apertures centered azimuthally on the fiducial midplane (see Section \ref{sect:midplane}) and centered radially at each pixel from the star. The apertures were [11$\times$4], [3$\times$3], [18$\times$18], and [30$\times$30] pixels in dimension (azimuthal $\times$ radial diameter). The first size was our best guess at an aperture that would capture the maximum amount of disk flux and minimum negative brightness in the inner part of the disk, where we are most interested in the structure. The other sizes were chosen to be much smaller and much larger than our optimal choice. For an aperture of a given size, we determined a correction factor to account for the likelihood that some disk brightness falls outside of the finite aperture. The disk brightness apertures are allowed to overlap in the radial direction, so no correction factor is needed for that dimension. Therefore, we construct an aperture with the same azimuthal width as that used for the disk brightness measurements but with an infinite radial width and compute the correction factor as the ratio of the flux within that aperture centered on an unocculted reference star to the total flux of the reference star. This correction factor was then divided into all surface brightness measurements for the given aperture.

To estimate the uncertainty on a disk surface brightness measurement at a given radial separation, we placed non-overlapping apertures in regions empty of disk brightness at the same radial separation, and took the standard deviation of the mean surface brightnesses within those apertures. We then added this standard deviation term in quadrature with a Poisson photon noise estimate to get the total uncertainty. The surface brightnesses and their associated uncertainties were converted to one-dimensional integrated brightnesses (mJy arcsec$^{-1}$) by integrating over the azimuthal dimension of the aperture for comparison with our model profiles.

As Figure \ref{fig:apBrights} shows, the four different aperture sizes produced significantly different brightness profiles in $H$, even after correcting for finite aperture size. All four profiles show a general trend of decreasing brightness with increasing separation, with shallower slopes in the inner disk and steeper slopes in the outer disk. Overall, larger apertures show shallower slopes than smaller apertures in both the inner and outer disk. In terms of absolute brightness, the ``optimal'' [11$\times$4] profile is brightest in the inner disk, the [30$\times$30] profile is dimmest, and the other profiles fall in between the two. For the outer disk, the [30$\times$30] profile is brightest, the [3$\times$3] profile is dimmest, and the other profiles fall in between the two. We do not plot the $K_s$ profiles but note that they show similar trends and discrepancies on the same order.

There are two effects related to aperture size that drive the differences between these profiles. First, smaller apertures fail to capture all of the flux from the disk, which biases the integrated brightness downward relative to larger apertures. This explains why the [3$\times$3] profile is consistently below the other profiles in the outer disk. Second, the larger apertures encompass more of the negative brightness created by self-subtraction on either side of the midplane, thereby biasing the integrated brightness downward relative to smaller apertures. This explains why the [18$\times$18] and [30$\times$30] profiles are dimmer than the [11$\times$4] profile in the inner disk, as the latter aperture was chosen specifically to match the width of the disk in this region and not include negative brightness. The outer disk exhibits less negative brightness, so the larger apertures do not suffer as much from this effect there and they tend to produce brighter profiles than smaller apertures. The two effects appear to balance each other at intermediate separations for the [3$\times$3] and [30$\times$30] apertures because their profiles show similar brightnesses at $r \approx115$ AU.

In comparison with the $H$ model profiles, the aperture profiles for all sizes are generally equal or lower in brightness for a given separation. The ratio of model brightness to the ``optimal'' $H$ [11$\times$4] aperture brightness is $\gtrsim 1$ for all $r$ (except at $r < 60$ AU, where the ratio is as low as $\sim0.7$ but the two profiles are statistically consistent), and is predominantly between 1.5 and 7, with a possible trend of the ratio increasing with separation. When we compare the $K_s$ model and aperture profiles, the model profiles are generally between 2 and 8 times brighter than the [11$\times$4] aperture profiles. Consequently, in both ansae of both bands, we find that our modeling process produces brightness profiles equivalent to or brighter than profiles produced by a best-case aperture. This is evidence that our algorithm recovers brightness lost to self-subtraction and, though we may be biased by the use of the spline model, it does so without the uncertainties involved with selecting an appropriate aperture size.

Another factor to consider when measuring brightness is ease of comparison between datasets and studies. One-dimensional integrated brightness profiles are more straightforward to compare across different wavelengths, datasets, and implementations than are aperture photometry profiles. This is because the one-dimensional profiles are integrated along the azimuthal dimension, leaving room for uncertainty only in the treatment of the measurement in the radial direction. This is one less than the two dimensions that can vary in standard aperture photometry measurements. Even in the absence of self-subtraction, this advantage remains. There is also an additional advantage of a relationship between the one-dimensional integrated brightness and the vertical optical depth to scattering presented by the grains \citep{graham2007}, which provides the opportunity to gain information about the latter quantity.

\subsection{Model Structure of HD 32297} \label{sect:structure}

Our HD 32297 observations benefit from some of the highest-angular-resolution imaging of the system to date. They are a valuable addition to the existing library of observations for this system, as high-contrast imaging is inherently difficult and results are dependent on the instrument and techniques used. As discussed in Section \ref{sect:brightness}, the various systematic errors involved with different types of observations and data analysis routines act as incentive to measure the properties of this disk using multiple independent datasets. Here we discuss several implications of our results for the physical structure of the debris disk.

\subsubsection{Brightness Profile Implications}

Brightness asymmetries in debris disks can act as signposts of planets or collisions within the disk (e.g., \citealt{wyatt1999, ozernoy2000}). However, we do not believe that the asymmetries we see in our radial brightness profiles indicate the existence of either in this system. Although some of the asymmetries appear statistically significant, this is still largely a proof-of-concept test and uncertainties in the systematics involved with high-contrast imaging and our disk modeling (such as artificially-inflated FWHM) preclude us from drawing further conclusions. Overall, the general symmetry of the disk ansae could be the result of a relatively smooth dust distribution or it could be due to the fact that when viewing an optically-thin disk edge-on, a given projected separation averages over the dust density at many radial separations.

The breaks in the brightness profile power laws we reported are perhaps more informative. Similar breaks in brightness profiles from previous studies of HD 32297 \citep{debes2009, boccaletti2012, currie2012_32297} have been attributed to the presence of a planetesimal ``birth ring'' located at the break radius \citep{strubbe2006}. Our measurements of the break point would locate the ring in the range of 95 AU $\lesssim r \lesssim 125$ AU. This is also consistent with the location of the cold dust ring estimated to be centered at 110 AU by \citet{donaldson2013}. Translation of our break location to a planetesimal ring radius requires modeling of the grain scattering phase function, which is beyond our current scope.

The differences between our ADI-corrected results and those from NICMOS observations are interesting because both methods have distinct associated systematic uncertainties. We have already discussed some of the potential systematic errors involved with our observations and data analysis. HST observations by \citet{schneider2005} and \citet{debes2009} also have potential systematic errors, though different from ours. For example, a spectral mismatch between the PSF reference star and HD 32297 could lead to over- or under-subtraction of the stellar PSF, as could PSF variations caused by the telescope's ``breathing'' \citep{fraquelli2004}. In addition, the thermal background from the telescope could bias the photometry, particularly at 2.05 $\mu$m, as \citet{debes2009} does not mention performing sky subtraction as recommended for observations with $\lambda\gtrsim1.7\ \mu$m \citep{nicmos2009}. Registration errors between stacked images are another possible error source in coronagraphic observations. Each of these systematic errors could impact the NICMOS photometry, and therefore affect asymmetries in radial brightness profiles or disk colors (see Section \ref{sect:color}). Our brightness profiles agree more closely with previous ground-based results than with the HST results and do not settle the question of which observational method provides more accurate information about disk. Instead, we suggest that further observations and more complete understanding of the errors associated with high-contrast imaging and related image-processing techniques are needed to answer this question with greater certainty.

\subsubsection{Implications of FWHM for Disk Shape and Self-subtraction Modeling} \label{sect:fwhm_discussion}

Our radial FWHM profiles show that the disk's projected width generally increases as $r$ increases. Evidence for projected width increasing with separation has been found in the roughly edge-on disks of $\beta$ Pictoris (\citealt{golimowski2006} and references therein), HD 15115 \citep{rodigas2012}, and AU Mic \citep{graham2007}. The indices of the best-fit power laws for our FWHM profiles were all between 1.27 and 1.47, which, for example, are slightly steeper than the indices ranging from 0.6 to 0.9 that \citet{kalas1995} found for the $\beta$ Pic disk at separations of $\sim$130$-$330 AU. One exception to the trend of increasing width in our data is at $r < 60$ AU, where we see a rise in both $H$ and $K_s$ widths as $r$ decreases. With residual speckle noise still contaminating the disk signal this close to the star, this feature is likely spurious. However, new observations or data reductions that provide a smaller inner working angle would allow us to investigate whether the disk continues to narrow closer to the star or whether some mechanism is causing the disk to ``puff up'' at separations less than 60 AU. Future extreme-AO instruments, such as GPI and SPHERE, promise to offer such capability and help answer these questions. Three-dimensional modeling of the disk would also allow us to ascertain the actual disk width rather than just the projected width.

Additionally, we do not see any sharp features in the FWHM profile that are significant. Such a feature could conceivably indicate the gravitational perturbation of grain or parent-body orbits, but our measurements do not clearly indicate such a scenario.

The wider FWHM that we see in $K_s$ compared to $H$ could be due to differently distributed dust populations responsible for scattering the two wavelengths or another physical mechanism. On the other hand, it could be a sign that our model overestimated the FWHM in $K_s$ or underestimated it in $H$. We apply the same modeling method to both datasets, so it would be unexpected for one set of measurements to be systematically offset due to something in the fitting process, yet we cannot rule this out because there may be systematics that are not fully understood. Consequently, we take the measurements at face value and refrain from further interpretation.

Comparison of our $K_s$ FWHM measurement with previous works may provide more support for the efficacy of our self-subtraction modeling algorithm. As noted in Section \ref{sect:fwhm}, our FWHM values for a given separation are generally smaller than the values reported by \citet{currie2012_32297} and greater than those reported by \citet{boccaletti2012}. This is telling because \citet{currie2012_32297} used a conservative LOCI ($N_\delta \geq 2.5$, W=5.5 AU) that would reduce self-subtraction while \citet{boccaletti2012} used a more aggressive LOCI (for the measurements in question; $N_\delta=1.0$, W=7.3 AU). We acknowledge that absolute comparisons between measurements from those works and our own are complicated by the different methods used. For example, the other works measured the FWHM of a Gaussian function fit to the vertical profile, while we fit a monotonic spline. Each work also included different corrections for self-subtraction in their measurements. Nevertheless, putting aside those issues and the complications of comparing reductions of different datasets with slightly different algorithms, it appears that our measurements fall between those taken from the aggressive and conservative reductions. This implies that our self-subtraction modeling algorithm is able to recover the disk brightness above and below the midplane that is removed by LOCI.

\subsubsection{Possible Sources of Disk Curvature} \label{sect:disc_curv}

Our midplane position results showed a northward curvature that qualitatively matched previous publications and that may be explained by existing hypotheses. One such hypothesis is that the HD 32297 system is interacting with the interstellar medium (ISM). This could cause brightness asymmetries \citep{kalas2005_32297} or warping of the outer disk by several degrees \citep{debes2009}. The estimated radial velocity of HD 32297 ($v\approx20\ \mathrm{km\ s^{-1}}$) is similar to that measured for the ISM ($v\approx24\ \mathrm{km\ s^{-1}}$; \citealt{redfield2007}), but HD 32297's proper motion is primarily to the south, which could potentially cause the disk's dust grains to be swept towards the north. However, this also depends on the proper motions of the ISM clouds and little is known about individual clouds in this region.

An alternative explanation is that a combination of scattering phase function and disk inclination can produce an observed curvature in the scattered-light surface brightness \citep{kalas1995}. If the disk is a few degrees from being perfectly edge-on and the dust grains are primarily forward scattering, then the near side of the disk (between us and the star) will appear brighter than the far side. This could account for the curvature we see toward the north if the near side of the disk is the south edge of the disk in our images. A similar explanation was proposed for the curvature observed in the HD 15115 debris disk \citep{rodigas2012}. \citet{currie2012_32297} modeled this scenario for HD 32297 and found that a disk containing highly forward scattering grains can cause a brightness asymmetry between the near and far sides that is more pronounced at small separations, causing a change in the midplane position angle and an apparent warping. \citet{boccaletti2012} also surmised that anisotropic scattering leads one edge of the disk to appear brighter than the other. Highly porous grains are expected to be forward-scattering \citep{graham2007} and might be one explanation for this feature, as \citet{donaldson2013} found that models of the outer disk that were composed of highly porous (90\% porosity) and icy grains provided the best fit to the disk's SED.

A complete explanation of the curvature may require both mechanisms. The difference in the amount of forward-scattered light observed from the near and far sides of the disk is least in the outer regions of the disk due to our viewing geometry, so the inclination may have little effect there. However, interaction of disk grains with ISM dust grains is expected to be stronger farther from the star \citep{artymowicz1997}, so an ISM interaction could produce curvature at the edges of the disk while inclination effects lead to curvature of the inner disk. We also note that the midplane curvature abruptly begins to increase just beyond the average break location of our brightness profile power laws, which may indicate that the same physical mechanism is responsible for both features.

\subsubsection{Implications of Disk Color for Grain Properties} \label{sect:color_discussion}

The color profiles derived from our one-dimensional (azimuthally-integrated) brightness measurements indicate a generally red color for the disk (Figure \ref{fig:modelColor}). This may imply grain sizes that are a few times larger than the approximately micron-sized grains typically thought to be the main source of scattering. \citet{donaldson2013} calculated the blowout size for the disk grains to be about 1 $\mu$m (assuming spherical, astrosilicate grains) and derived a minimum grain size of 2.1 $\mu$m from their best-fit SED model. Both of these values would be roughly consistent with the grain sizes implied by our color estimate.

Alternatively, one can compute the disk color based on peak brightness rather than integrated brightness. Our peak brightness measurement is also intrinsically corrected for self-subtraction, although self-subtraction is typically least influential near the midplane, where we assume the peak to occur. We find the peak brightness profiles for both ansae to be slightly brighter in $H$ than $K_s$, leading to a peak color that is slightly blue at most separations and gray at the rest. This agrees with the colors found by \citet{debes2009} and is consistent with the disk being populated by micron- or submicron-sized grains that scatter 1.6 $\mu$m and 2.15 $\mu$m light with approximately the same efficiency. 

The discrepancy between the two methods is possibly due to imprecision in the modeled disk FWHM measurement. As referred to in Section \ref{sect:fwhm}, the $K_s$ FWHM is generally larger than the $H$ FWHM at a given separation. This greater width increases the $K_s$ integrated brightness relative to the $H$ integrated brightness but does not affect the peak brightnesses. Thus, the disk may appear red from one interpretation and blue/gray in the other. This is a case where a more complex model of the disk vertical profile would allow us to gain a more precise understanding of the dust population. The color derived from the peak brightness is likely more reliable because it is less sensitive to the model choice than is the integrated brightness color, but this has the disadvantage of only probing the dust at the midplane.

\section{CONCLUSIONS} \label{sect:conclusions}

We have presented a novel technique for forward-modeling self-subtraction in ADI/LOCI-processed images of extended emission and applied it to near-infrared scattered-light imaging of the HD 32297 debris disk. Our method successfully reproduced the self-subtraction pattern in our $H$- and $K_s$-band LOCI images using a relatively simple model of the disk's vertical brightness profile, the LOCI parameters, and the position angles of the images in the ADI dataset. The result of the modeling process was a model of the disk's two-dimensional surface brightness that was not distorted by self-subtraction and provided an approximation of the scattered-light distribution prior to image processing. In the future, this self-subtraction modeling could be used in combination with other versions of ADI or LOCI, such as those that use iterative reference PSF subtractions, masks, or damped coefficients.

From the self-subtraction-corrected models we extracted radial profiles for the one-dimensional integrated brightness, projected FWHM, and midplane location of the disk. The brightness profiles did not indicate any clear asymmetries or structures but our power-law fits showed a break that supports the existence of a planetesimal birth ring at $r\approx110$ AU. We also demonstrated that the model-derived profiles contained less uncertainty from location-dependent self-subtraction and variable disk width than profiles measured via aperture photometry. The FWHM profiles indicated a projected disk width that increased with separation from the star. Our measurements of midplane location showed curvature towards the northwest that confirms previous reports of similar features. This curvature may be a combination of a geometric observational effect linked to the disk's nearly edge-on inclination and interaction of the disk with the ISM. Additionally, we found the disk's color to depend on our choice of model for the disk vertical profile but we estimate the midplane color to be blue or gray at all separations.

Our self-subtraction modeling technique provides a two-dimensional model of the disk's scattered light that is a good starting point for building a three-dimensional model that can provide more information about grain size, grain composition, dust density, and disk morphology. The speed and accuracy with which we can  compute the self-subtraction function using this method will also be extremely useful in future work that will compare more-complex two-dimensional model images with ADI-processed observations. We hope to apply our technique to a more detailed investigation of this system and to ground-based high-contrast AO observations of other circumstellar disks with the goal of learning more about the origins of planetary systems.

\acknowledgments

The authors wish to thank the anonymous referee for helpful suggestions that improved this manuscript. This work was supported in part by University of California Lab Research Program 09-LR-01-118057-GRAJ, NSF grant AST-0909188, and NASA Origins grant 09-SSO09-0124. T. Esposito was supported by a UCLA graduate research fellowship for the duration of this work.
\par The data presented herein were obtained at the W. M. Keck Observatory, which is operated as a scientific partnership among the California Institute of Technology, the University of California, and the National Aeronautics and Space Administration. The Observatory was made possible by the generous financial support of the W. M. Keck Foundation. The authors wish to recognize and acknowledge the very significant cultural role and reverence that the summit of Mauna Kea has always had within the indigenous Hawaiian community. We are most fortunate to have the opportunity to conduct observations from this mountain.

{\it Facility:} \facility{Keck:II (NIRC2)}

\clearpage

\bibliographystyle{hapj}   %>>>> makes bibtex use spiebib.bst
\bibliography{apj-jour,selfsub_refs}

\end{document}